\RequirePackage{fix-cm}
\documentclass[smallextended]{svjour3}       
\smartqed  
\usepackage{graphicx}
\usepackage{amsmath,amssymb}
\usepackage{epstopdf,epsfig}
\usepackage{pifont,bbding}
\usepackage{rotating}
\usepackage{hyperref}
\usepackage{natbib}
\usepackage{lscape}
\usepackage{mathptmx}      
     \newcommand\ergshz{erg~s$^{-1}$~Hz$^{-1}$}

     \def\Spitzer{\sc Spitzer}

\begin{document}

\input psfig.sty

\title{Atmospheres of Brown Dwarfs\thanks{ChH highlights financial
    support of the European Community under the FP7 by an ERC starting
    grant. SLC acknowledges financial support of University of
    Leicester. V. Wild is thanked for 'large-scale metallicity
    gradients'. M. Marley, A. Scholz, I. Vorgul, P. Rimmer,
    T. Robinson, I. Leonhardt are thanked for reading the manuscript. The authors
    thank D. Saumon, M. Cushing and J.D. Kirkpatrick for kindly
    providing spectra. This research has benefited from the SpeX
    Prism Spectral Libraries, maintained by Adam Burgasser and the
    IRTF spectral library maintained by Mike Cushing. Most literature
    search was performed using the NASA Astrophysics Data System ADS.
    Our local computer support is highly acknowledged.} }


\author{Christiane Helling         \and
        Sarah Casewell 
}


\institute{Ch. Helling \at
              SUPA, University of St Andrews, North Haugh, St Andrews, KY16 9SS, UK \\
              Tel.: +123-45-678910\\
              Fax: +123-45-678910\\
              \email{ch80@st-and.ac.uk}           
           \and
           S. Casewell \at
              Department of Physics and Astronomy, University of Leicester, University road, Leicester, LE1 7RH
}

\date{[Received:] Version: 26 September 2014 / Accepted: 22 October 2014}

\maketitle

\begin{abstract}
Brown Dwarfs are the coolest class of stellar objects known to date.
Our present perception is that Brown Dwarfs follow the principles of
star formation, and that Brown Dwarfs share many characteristics with
planets. Being the darkest and lowest mass stars known makes Brown
Dwarfs also the coolest stars known. This has profound implication for
their spectral fingerprints. Brown Dwarfs cover a range of effective
temperatures which cause brown dwarfs atmospheres to be a sequence
that gradually changes from a M-dwarf-like spectrum into a planet-like
spectrum. This further implies that below an effective temperature of
$\lesssim 2800$K, clouds form already in atmospheres of objects
marking the boundary between M-Dwarfs and brown dwarfs.  Recent
developments have sparked the interest in plasma processes in such
very cool atmospheres: sporadic and quiescent radio emission has been
observed in combination with decaying Xray-activity indicators across
the fully convective boundary.
\end{abstract}

\section{Introduction}

\begin{figure}
\centering
\includegraphics[width=\textwidth]{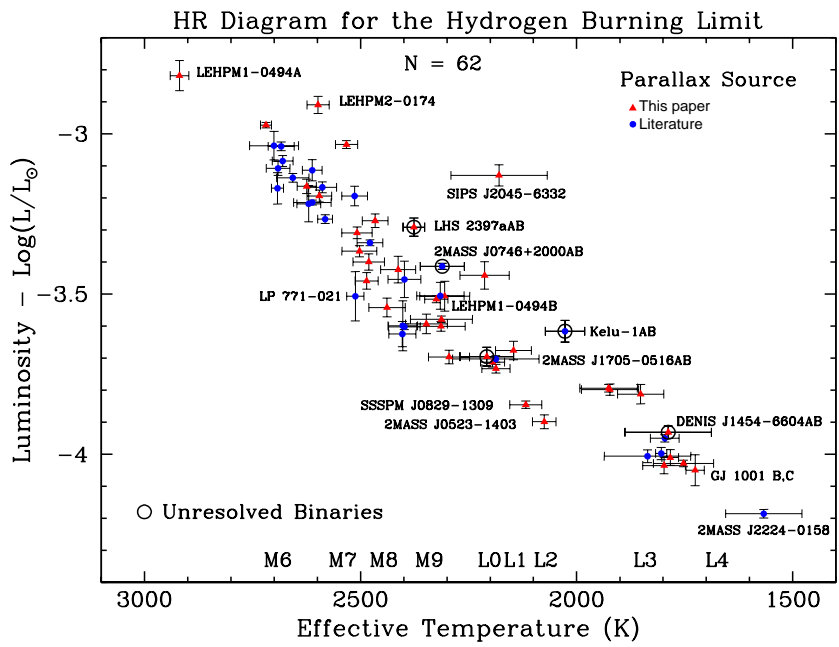}
\caption{The low-mass end of the main sequence in the HRD diagram
  showing the transition from the stellar M-dwarf to the substellar
  brown dwarf regime (\citealt{diet2013}; courtesy: S. Dieterich).}
\label{diet13_HRD}
\end{figure}

Brown dwarfs are free-floating stellar objects with masses below the
hydrogen-burning limit of $\sim 0.075$M$_{\odot}$ and with radii of
$\approx 1 R_{\rm Jup}$ for mature objects, although this can vary
with cloud cover and metallicity. They are the low-mass extension of
the sub-solar main-sequence in the Hertzsprung-Russel diagram (HRD,
Fig.~\ref{diet13_HRD}).  Their total emitted flux, and hence their
effective temperature T$_{\rm eff}$, is lower than that of
M-dwarfs. Brown dwarfs become increasingly dimmer as they age because
their mass is too small to sustain continuous hydrogen burning. Only
in their youth, the heaviest brown dwarfs fuse some helium and maybe
lithium. Hence, the class of brown dwarfs comprises members that are
just a little cooler than M-dwarfs (L dwarfs) and members that can be
as cold as planets (T and Y dwarfs). Several formation mechanisms are
suggested, including the classical star-forming scenario of a local
gravitational collapse of an interstellar molecular cloud. The
formation efficiency may have changed depending on time and
location. The oldest brown dwarfs could be as old as the first
generation of stars that formed in the universe. Their metallicity
would be extremely low leaving the spectrum almost feature-less
(Fig. 9 in \citealt{witte2009}). \cite{luh2012} reviews the formation
and evolution of brown dwarfs, including the initial mass function and
circumstellar disks. Observational evidence builds up for that brown
dwarfs and giant gas planets overlap in masses and in global
temperatures (see review \citealt{chab2014}). \cite{viki14} reviews the
emergence of brown dwarfs as a research area started by a theoretical
prediction of their existence, and emphasizes the research progress in
formation and evolution of brown dwarfs.
\cite{allard97} reviewed model atmospheres of very low mass stars and
brown dwarfs discussing model aspects like updating gas-phase
opacities, convection modelled by mixing length theory, and the
T$_{\rm eff}$-scale for M-dwarfs. The present review summarizes the
progress in brown dwarf observations extending now from the UV into
the radio, revealing new atmospheric processes and indicating
overlapping para\-meter ranges between brown dwarfs and planets
(Sect.~\ref{s:obs}). Special emphasis is given to cloud modelling as
part of the brown dwarf atmospheres, a field with increasing
importance since the first review on brown dwarf atmospheres by
\cite{allard97} (Sect.~\ref{s:theo}). Since \cite{allard97}, a
considerably larger number of brown dwarfs is known which allows first
statistical evaluations, one example being the search for correlation
between X-ray emission and rotational activity in brown
dwarfs. Wavelength-dependent variability studies have gained momentum,
and the idea of weather on brown dwarfs is commonly accepted since the
first variability search by \cite{tt99}. In the following, we
summarize the observational achievements for field brown dwarfs. We
ignore the specifics of young brown dwarfs as this is reviewed in
e.g. \cite{luh2012} and concentrate on evolved brown dwarfs, not on
individual star forming regions or brown dwarfs with
disks. Sections~\ref{s:bench} discusses the idea of model benchmarking
as it emerged in the brown dwarf community. Section~\ref{s:compl}
gives an outlook regarding new challenges like multi-dimensional
atmosphere modelling and kinetic gas-phase chemistry.

\section{Brown dwarf observations in different spectral energy ranges}\label{s:obs}
In this section of the review we aim to bring the reader up to date
with the observations of brown dwarfs. We describe the recent results
focusing on cloud effects in low gravity objects, and their comparison
with young extrasolar planets, as well as observations of low
metallicity objects.  We also discuss the recent reports of
photometric and spectroscopic variability of brown dwarfs, which is
linked to patchy clouds, temperature fluctuations within the
atmosphere and weather effects, as well as high energy phenomena that
are linked to emissions seen in the radio, X-ray and UV wavelength
regimes.

\subsection{Optical and IR spectral types}\label{ssOIR}

Although the first brown dwarf to be discovered was the L4 dwarf
GD165B \citep{becklin88}, it was not identified as such until the
discovery of Gl229B \citep{nakajima95} and Teide 1 \citep{rebolo95} in
1995. Since then, astronomers have searched for a way of classifying
brown dwarfs. These objects are very different from the M dwarfs
known at the time, and the contrast between the dusty L dwarfs and the
methane rich T dwarfs is stark. Both these spectral types are
discussed extensively in the literature and are not described here
in great detail (\citealt{burrows01, lodders06} for an overview).

\begin{landscape}
\begin{figure}
\begin{center}
{\ }\\*[-1cm]
\hspace*{-0cm}
\includegraphics[width=0.83\textwidth]{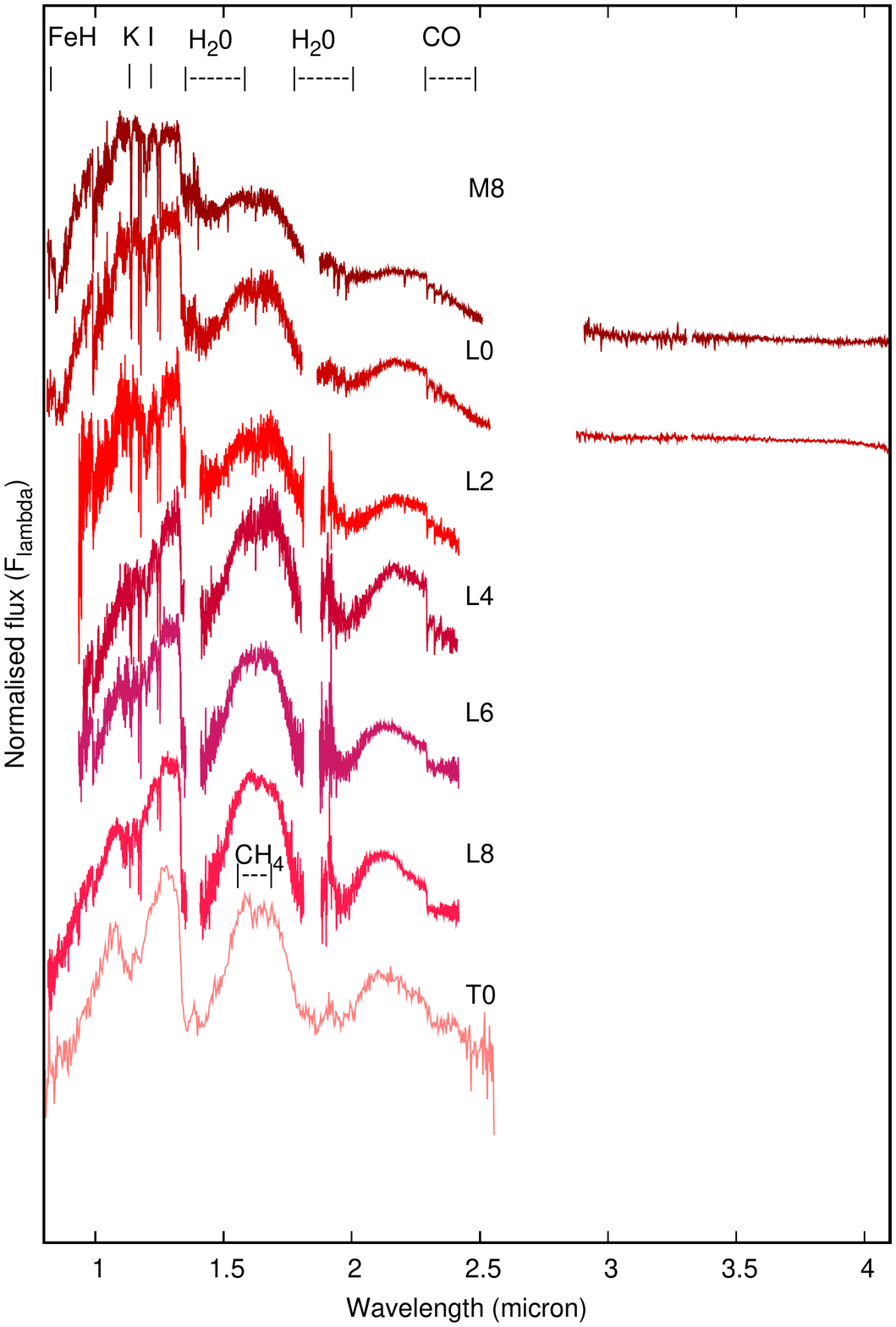}
\hspace*{-1.3cm}
\includegraphics[width=0.83\textwidth]{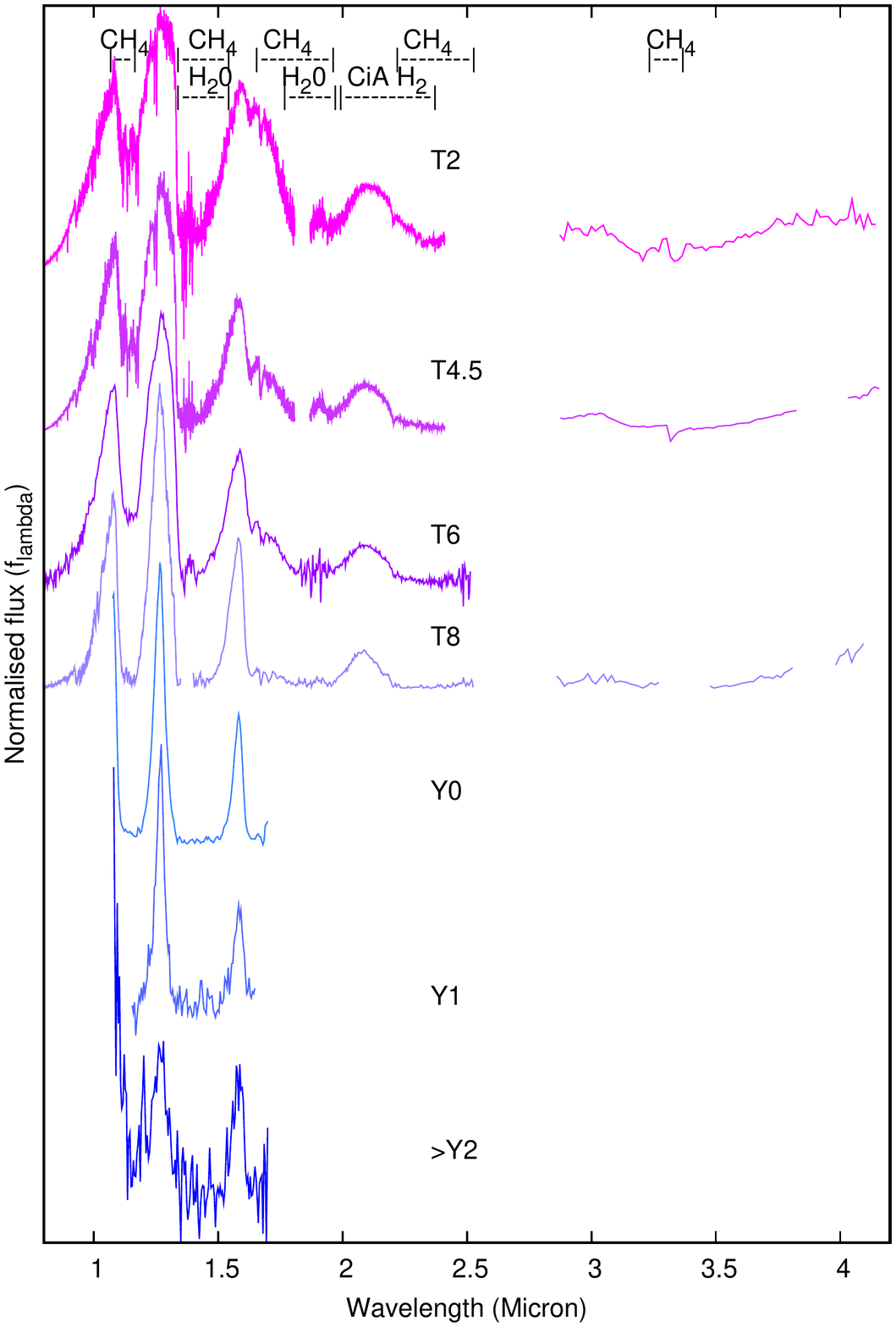}\\*[-0.8cm]
\caption{\label{st} 
The  near-infrared spectral sequence from early-M dwarfs to early-Y brown
  dwarfs. All spectra are normalised to 1 at their peak and then a
  flux offset is applied \citep{kirk13}. }
\end{center}
\end{figure}
\end{landscape}

The L dwarfs are similar to M dwarfs in photospheric chemical
composition, containing alkali lines of (K, Na) and metal hydrides
(FeH) and oxides (TiO, VO) and water.  As we progress through the
spectral types from L0 to L9, the TiO and VO bands weaken, the alkali
lines become weaker and more pressure broadened, and the water bands
and FeH strengthen in the optical. In the near-IR, CO strengthens
towards the mid-L dwarfs, and then weakens again as methane begins to
form.  The change between the L and T dwarfs, often called the L-T
transition region is characterized by the near-infrared colours of the
brown dwarfs changing, while the effective temperature of the objects
remains the same (Figure \ref{st}; see \citealt{kirkpatrick99} for a
review).
T dwarfs, sometimes called methane dwarfs are characterized by the
methane absorption seen in the near-IR that gets progressively
stronger as one progresses through the subclasses, making the $J-H$
and $H-K$ colours bluer.  In the optical, the spectrum is affected by
collisionally induced molecular hydrogen absorption and FeH.

2011, marked the discovery of an additional, and later spectral type,
the Y dwarf.  There are $\sim$20 Y dwarfs known to date
\citep{cushing11, luhman11, kirkpatrick12, kirkpatrick13,
  cushing14}. The majority of these have spectral types ranging
between Y0 and Y2 and were discovered using the \textit{Wide-field
  Infrared Survey Explorer} (WISE). WISE was designed to discover Y
dwarfs: its shortest wavelength band at 3.4$\mu$m was selected to fall
in the centre of the fundamental methane absorption band at 3.3$\mu$m
and the W2 filter at 4.6$\mu$m detects radiation from the deeper,
hotter layers in the atmosphere. When combined, the W1-W2 colour is
very large, allowing the detection of Y dwarfs \citep{kirkpatrick11}.

All the Y dwarfs show deep H$_{2}$O and CH$_{4}$ absorption bands in
their near-infrared spectra, similar to late-T dwarfs (Figure
\ref{st}). These water clouds were measured in detail by
\citet{morley14a}.  The $J$ band peaks of the Y dwarfs are narrower
than those of the latest type T dwarfs, and the ratio of the $J$ and
$H$ band flux is close to 1, meaning that the $J-H$ trend towards the
blue for T dwarfs turns back towards the red for Y0. This trend also
occurs for the $Y-J$ colour.  This colour reversal is thought to be
caused by alkali atoms, that normally dominate the absorption in the
shorter wavelengths, being bound in molecules, thus reducing the
alkali atom opacity \citep{liu10}. While the $H$ band spectra of T
dwarfs are shaped by CH$_{4}$ and H$_{2}$O, for Y dwarfs as the
effective temperature decreases, NH$_{3}$ absorption becomes important
\citep{lodders02, burrows03}.

\citet{cushing11} estimate T$_{\rm eff}=350\,\ldots\,500$K for the Y0
dwarfs, with their masses between $\sim$5 and 20 M$_{\rm Jup}$.
However, using a luminosity measurement derived from a $\Spitzer$
based parallax, \citet{dupuy13} estimate their effective temperatures
to be typically 60-90 K hotter. This difference in temperature is
possibly caused by using near-IR spectra, a regime where only 5\% of
the Y dwarf flux is emitted and so models do not always accurately
reproduce the observations \citep{dupuy13}.
Using models containing sulphide and chloride clouds from \cite{morley12},
\citet{beichman14} obtain the lower effective temperatures, suggesting
that previous results apply on a model dependent bolometric
correction. There is one other Y dwarf, not discovered by WISE:
WD0806-661B, which was until recently, the most likely candidate for
the lowest mass and temperature Y dwarf at 6-10 M$_{\rm Jup}$ and
330-375 K, although there are as yet no spectra of this object. This
has since been superseded by the discovery of WISEJ085510.83-071442.5,
a high proper motion Y dwarf at 2 pc. This object is our fourth
nearest neighbour and has an estimated effective temperature of
225-260 K and a mass of 3-10 M$_{\rm Jup}$ \citep{luhman14}.


\subsubsection{Brown dwarf classification}
Brown dwarf spectral types do not fit into the standard Morgan-Keenan
stellar spectral types system because their age-mass-temperature
degeneracy causes  the classical stellar mass-temperature
relationship to break down for such ultra-cool objects. An object of a
specific spectral type (or effective temperature) may be higher mass
and old, or lower mass and young. For instance, an L5 dwarf (T$_{\rm
  eff}$$\sim$ 1500 K) in the Pleiades cluster (125 Myr) has a mass of
25 M$_{\rm Jup}$, but a field dwarf of the same spectral type is much
more massive at 70 M$_{\rm Jup}$ \citep{chabrier00}.

There are in general three methods of identifying a brown dwarf's
spectral type: the first, by comparison with spectral templates as is
usual for stars that fit the Morgan-Keenan classification scheme, the
second is by using indices derived from spectral parameters and the
third is by comparing broadband photometry to spectral standards. The
first two methods are described for the L dwarf classification in the
optical wavelength range by \citet{kirkpatrick99} and \citet{martin99},
respectively.  The most commonly used method is to compare spectra of
objects to "standard" or "template" spectra and to use spectral
indices as a secondary calibration tool, for instance to judge
metallicity or gravity (see Sect.~\ref{grav} for the gravity
classification scheme). The template scheme was extended into the
near-IR by \citet{reid01}, and the indices by \citet{geballe02}. This
index scheme is now more widely used than that of \citet{martin99} and
extends down to T9. \citet{burgasser06a} combined the index scheme of
\citet{geballe02} with their template \citep{burgasser02} to create a
unified way of spectral typing T dwarfs. When using these methods to
classify L and T dwarfs it should be noted that the assigned spectral
types are limited to only the parts of the spectrum that are
measured. For objects on the L-T transition, it is not unusual to have
optical and near-IR derived spectral types that differ by $>$1
spectral type.

\subsubsection{Low metallicity brown dwarfs}
Low metallicity brown dwarfs (subdwarfs) provide an insight into the
coolest, oldest brown dwarfs. The existence of low-metallicity brown
dwarfs indicates that such low-mass stars also formed in a younger
universe when the metallicity was lower than today.  It is further of
interest to comparing their atmospheres to those of low metallicity
extrasolar planets which formed as by-product of star formation.

Only a handful of ultracool subdwarfs are known to date
\citep{burgasser03, burgasser04, cushing09, sivarani09, lodieu10,
  kirkpatrick10, lodieu12, mace13, kirkpatrick14, burningham14}. Of
the $\sim$30 objects known to date, only 11 have spectral types later
than L2 \citep{kirkpatrick14, burningham14, mace13}. The naming scheme
for subdwarfs follows that for M dwarfs developed by \citet{gizis97}
and upgraded by \citep{lepine}, moving from dM for metal rich M dwarfs
to subdwarfs (sdM), extreme subdwarfs (esdM) and ultra subdwarfs
(usdM) in order of decreasing metallicity.

\begin{figure}
\begin{center}
\scalebox{0.25}{\includegraphics[angle=0]{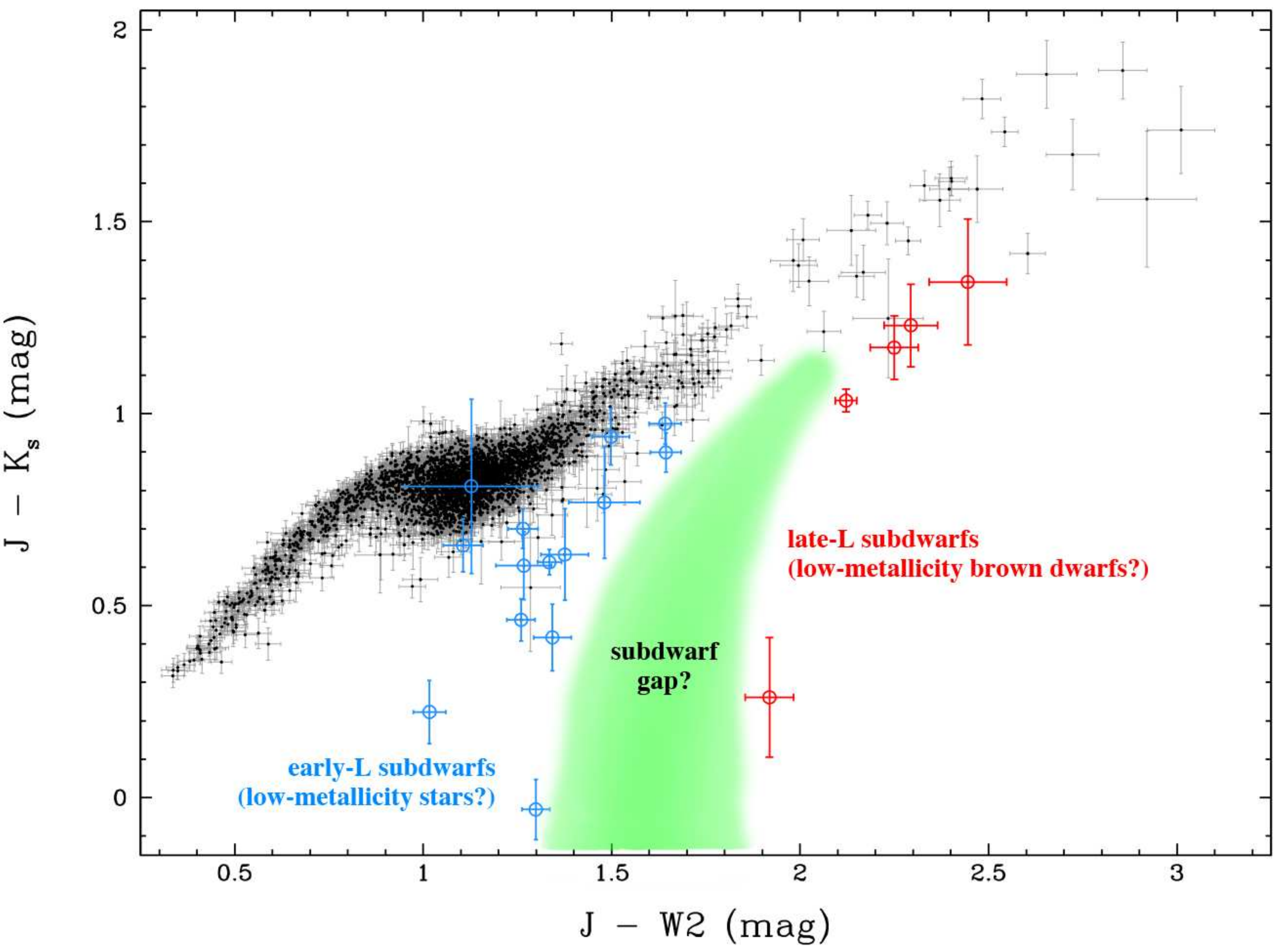}}
\caption{\label{sd} 
The
  effective temperature gap between brown dwarfs and M-dwarfs
  increases with lower metallicity (lower J-K$_{\rm s}$).  Known
  late-M (SpecT $<$ sdL5 -- blue circles) and L subdwarfs (SpecT $>$
  sdL5 -- red circles) and AllWISE proper motion stars (solid black
  dots) are shown (\citealt{kirkpatrick14}). 
  \cite{kirkpatrick14} suggest that the wedge (green zone) covers an
  area in the diagram where L subdwarfs may rarely be found. }
\end{center}
\end{figure}

 \citet{burgasser08} noted that L subdwarfs are overluminous in
 $M_{J}$, but slightly underluminous in $M_{K}$. This change is
 suggested to be caused by a reduced cloud opacity causing strong TiO,
 FeH, Ca $\textsc{I}$ and Ti $\textsc{I}$ features, and enhanced
 collisional-induced H$_{2}$ opacity in the $K$ band as predicted by
 \citet{ackerman01, tsuji96}.  These sources of opacity will be
 discussed in more detail in Section
 \ref{s:theo}. \cite{kirkpatrick14} suggests that an L subdwarf "gap"
 exists between the early L and late L subdwarfs. This could be
 explained if lower metallicity brown dwarfs would generally be the
 older objects.  For older brown dwarfs, an effective temperature gap
 is observed to occur between the hotter, deuterium-burning subdwarfs
 (blue symbols in Fig.~\ref{sd}), and the coolest, lowest mass members
 of similarly aged hydrogen burning stars (M-dwarfs, red symbols in
 Fig.~\ref{sd}). This temperature gap is predicted to increase for
 older populations \citet{kirkpatrick14}.

Recently, a population of T subdwarfs has emerged \citep{burningham14,
  mace13}. T dwarf colours are very sensitive to small changes in
metallicity - a shift of 0.3 dex can change the $H-$[4.5] colours of
T8 dwarfs as much as a 100 K change in effective temperature
\citep{burningham13}. The T dwarfs exhibit enhanced $Y$ and depressed
$K$ band flux, indicative of a high gravity, and hence older age, and
low metallicity atmosphere \citep{burgasser02}. The increased $Y$ band
flux is caused as the lower metallicity reduces the opacity in the
wings of the alkali lines, resulting in a brighter and broader peak
flux \citep{burgasser06}.  The $K$ band flux depression is created as
pressure-enhanced collision induced absorption by molecular hydrogen
becomes more important, as molecular features are removed from the
spectra \citep{saumon94}.

\subsubsection{The surface gravity of brown dwarfs}\label{grav}

Young, low gravity brown dwarfs have very similar properties to
directly-imaged exo\-planets \citep{fah2013} and it has been suggested
that younger brown dwarfs (log($g$)$\approx 3$) may have thicker
clouds in their atmospheres than those present in older objects
(log($g$)$\approx 5$) of the same effective temperature
\citep{barman11, currie11, madhusudhan11}.  \citet{hell2011}
demonstrate that the geometric cloud extension\footnote{The
  geometrical cloud extension, or cloud height, can be defined in
  various ways. \cite{woi2004} used the degree of condensation for Ti
  (their Eq. 16) for defining the cloud height.  The distance between
  the gas pressure at the nucleation maximum and the gas pressure
  where all cloud particles have evaporated determine the cloud height
  in \cite{hell2011}.} increases with decreasing surface gravity
log($g$) in cloud-forming atmospheres, an effect likened to an
increasing pressure scale height H$_{\rm P}\sim$1/$g$, and
\citet{marley2012} suggest that these clouds persist for longer, at
higher temperatures, than in older objects.

Many brown dwarfs have been identified in young open star clusters
(e.g. Sigma Ori: \citep{bihain09, pena12}, Serpens: \citep{spezzi12};
Pleiades: \citep{casewell11, casewell07}; see \citealt{luhman12} for a
review). However, there also exists a field population of low gravity,
young brown dwarfs (e.g. \citealt{reid08, cruz09, kirkpatrick10}).  A
comprehensive scheme for defining the gravity of these objects was
devised by \citet{kirkpatrick08} and \citet{cruz09} in the optical,
and \citet{allers13a, allers13b} in the near-infrared. The
classification introduces a suffix of $\alpha$, $\beta$, or $\gamma$
to the spectral type, indicating the gravity.  $\alpha$ implies a
normal gravity field dwarf, whereas $\beta$ is an intermediate-gravity
object and $\gamma$ represents a low-gravity object. These suffixes
can also be used as a proxy for age.

In the optical, the suffixes are assigned based on measurements of the
Na $\textsc{I}$ doublet and the K $\textsc{I}$ doublet which are
weaker and sharper in a low gravity object, the VO absorption bands
which are stronger, and the FeH absorption features which are weaker.
We note that decreasing hydride molecules are typical for low
metallicity stars of higher mass, hence higher temperature. The
abundance of such hydrogen-binding molecules decreases in brown dwarfs
because cloud formation decreases the metal components available.

 In the near-infrared, the VO and FeH bands are considered
 simultaneously with the alkali lines to derive $\log$(g). The
 changing shape of the $H$-band is also taken into consideration.  It
 becomes more triangular-shaped caused by increasing water absorption
 as a sign of low gravity \citep{lucas01, rice11}. All of these
 features are altered as there is less pressure broadening due to the
 object having a low surface gravity causing the pressure scale height
 to increase \citep{rice11, allers13a}.  In general, alkali absorption
 features are weaker, and the overall colours of lower gravity objects
 are redder than those of their higher gravity counterparts
 \citep{faherty13}. The redder colour is due, in part, to the changes
 in the near-infrared broadband features, but also due to more
 photospheric dust \citep{hell2011}. An additional feature is that
 while young M dwarfs ($\log$(g)$\approx 3$) are brighter than their
 older (log($g$)$\approx 4.5$) counterparts, young L dwarfs
 (log($g$)$\approx 3$) are underluminous in the near-infrared for
 their spectral type. This may be due to the additional dust in their
 photosphere \citep{faherty13a} or due to a cooler spectral
 type/temperature relation being required.

There are a handful of directly-imaged planetary mass companions that
have estimated effective temperatures in the brown dwarf regime
(e.g. \citealt{bonnefoy13}), for example 2M1207b ($\sim$ 10 Myr,
$\sim$ 1600 K, SpecT$\sim$mid L) and HR8799b ($\sim$30 Myr, $\sim$1600
K, SpecT$\sim$ early T). These planets are underluminous and have
unusually red near-infrared colours, when compared to field brown
dwarfs, as well as displaying the characteristic peaked $H$-band
spectra. There are $\sim$30 brown dwarfs that have been kinematically
linked to moving groups and association aged between 10 and 150 Myr,
that share these features, thus indicating that low gravity brown
dwarfs may provide a clue to the atmospheric processes occurring on
young exoplanets.

\subsection{High energy processes in non-accreting brown dwarfs}

\begin{figure}
\begin{center}
\scalebox{0.82}{\includegraphics[angle=0]{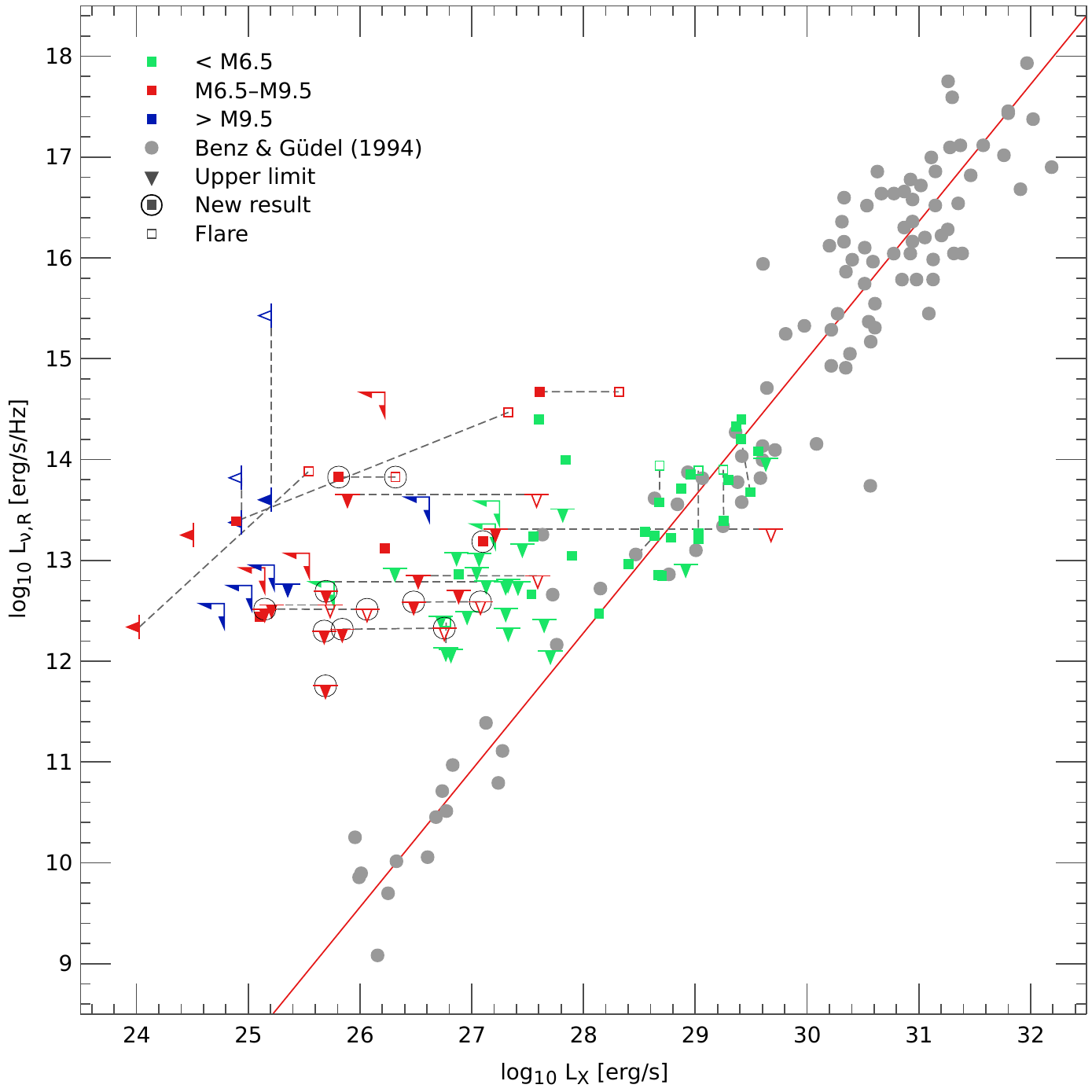}}
\caption{\label{gb} 
The G{\"u}del-Benz relationship between $L_{X}$ (0.2-2 keV) and
$L_{v,R}$ (\citealt{williams13c}). Limits are shown as downward
pointing triangles. Objects with spectral types of M6 or earlier
(green), M6.5-M9.5 dwarfs (red) and spectral types of L0 or later
(blue) are shown. Grey circles show the original G{\"u}del-Benz
relation from \citet{benz94}. Dashed lines connect multiple
measurements of the same source.}
\end{center}
\end{figure}

Despite brown dwarfs being objects that are brightest in the near-IR,
this has not prevented searches for other types of emission,
particularly those associated with higher energy processes such as
those seen in early M dwarfs.  However, the magnetic dynamo mechanism
is used to explain magnetic field generation in solar-type stars, and
there is a direct correlation between rotation and magnetic activity
indicated by H$\alpha$, X-rays and radio emission \citep{noyes84,
  stewart88, james00, delfosse98, pizzolato03, browning10, reiners09,
  morin10}. This relationship between the radio (L$_{v,R}$) and X-ray
(L$_{X}$) luminosities holds for active F-M stars, and L$_{X} \propto
L_{v,R}^{\alpha}$ ($\alpha$$\sim$0.73) is known as the G{\"u}del-Benz
relation (Figure \ref{gb}: \citealt{gudel93, benz94}).  As the dynamo
operates at the transition layer between the radiative and convective
zones (the tachocline), this mechanism cannot explain radio activity
in fully convective dwarfs ($>$M3), and although H$\alpha$ and X-ray
activity continues into the late M dwarf regime, the tight correlation
between X-ray and radio luminosity breaks down which suggests that a separate
mechanism is likely to be responsible for radio emission.

\subsubsection{X-ray and UV observations}\label{X-ray}

Many searches for X-ray emission in L dwarfs were conducted \citep{stelzer03, berger05, stelzer06}, but only one detection is known to date.  The L dwarf binary Kelu-1, composed of two old brown dwarfs, was detected with $Chandra$ in the energy range of 0.63, 0.86, 1.19 and 1.38 keV, resulting in an estimation of the  0.1-10 keV X-ray luminosity to be $L_{X}=2.9\times10^{25}$ erg s$^{-1}$ \citep{audard07}. It has been suggested that this emission does not originate from flares as there was no concurrent radio detection at a frequency $\sim$8 GHz \citep{audard07}.
\cite{audard07} suggested that the ratio between luminosity (radio, H$\alpha$, or X-rays) and bolometric luminosity (L/L$_{\rm bol}$) 
 increases with decreasing effective temperature.  They concluded that the chromospheric magnetic activity (H$\alpha$ emission) and the activity in the hot coronal loops (X-ray emission) decreases with effective temperature, indicating a different mechanism responsible for the radio emission in ultra cool dwarfs.

 \citet{williams13c} suggest that ultracool dwarfs with strong axisymmetric magnetic fields tend to have $L_{v,Radio}/L_{X}$ consistent with the G\"{u}del-Benz relation (Fig.~\ref{gb}),  while dwarfs with weak non-axisymmetric fields are radio luminous. Slower rotating dwarfs have strong convective field dynamos and so also stay near the G\"{u}del-Benz relation, whereas some rapid rotators may violate the G\"{u}del-Benz relation. \citet{williams13c} also note that dwarfs with a weaker magnetic field  tend to have later spectral types, and lower X-ray luminosity, which may be related to their cooler temperatures. They also note that in general, radio-bright sources, tend to be X-ray under luminous compared to  radio dim brown dwarf sources.

There are as yet no UV detections of brown dwarfs where the emission
is attributed to atmospheric processes.  Only brown dwarfs with disks
such as the young TW Hydra member 2MASS1207334-39 have been observed
to show UV emission. H$_{2}$ fluorescence is detected and attributed
to accretion \citep{gizis05}.

\subsubsection{Optical H$\alpha$ emission}\label{ss:halfa}

The H$\alpha$ luminosity (indicative of chromospheric activity) in
late M and L dwarfs also decreases with lower mass and later spectral
type. \citet{schmidt07} estimate that of their 152 strong sample, 95\%
of M7 dwarfs show H$\alpha$ emission (consistent with
\citealt{gizis00}). This fraction decreases with spectral type to 8\%
of L2-L8 dwarfs showing H$\alpha$. For the L dwarfs in particular,
50\% of L0 dwarfs being active declines to 20-40\% of L1 dwarfs and
only 10\% of L2 dwarfs and later spectral types. This decline is
similar to the breakdown in the rotation-activity relationship seen
for the X-ray activity (Section \ref{X-ray}) and has been attributed
to the high electrical resistivities in the cool, neutral atmospheres
of these dwarfs \citep{mohanty02}.

\citet{sorahana} have recently suggested that molecules may be
affected by chromospheric activity. The active chromosphere heats the
upper atmosphere, causing the chemistry in that region to change,
resulting in the weakening of the 2.7 $\mu$m water, 3.3$\mu$m methane
and 4.6$\mu$m CO absorption bands as seen in $\textit{AKARI}$ spectra
of mid-L dwarfs.

Despite the predicted decrease in H$\alpha$ emission, some objects of
late spectral type show H$\alpha$ emission. For example, the L5 dwarfs
2MASSJ01443536-07 and 2MASS1315-26. In quiescence, these objects
show H$\alpha$ fluxes similar to those of other dwarfs of the same
spectral type. However in outburst, 2MASS0144-07 has an H$\alpha$
flux measurement more than 10 times higher than the mean
\citep{liebert03}, and for 2MASS1315-26 the H$\alpha$ emission is
$\sim$ 100 times stronger than for L dwarfs of a similar spectral type
\citep{hall02}.

\subsubsection{Radio emission}

A number of brown dwarfs have been detected to be radio loud.  This
non-thermal emission may be low level quiescent \citep{berger02},
exhibiting variability \citep{antonova07}, showing variability that is
periodic and linked to rotation \citep{berger05, mclean11}, highly
polarised and bursting \citep{burgasser05}, pulsing synchronised with
the rotation period, and highly polarised \citep{hallinan07, berger09}
or a combination of some of these \citep{williams13b}.  Some of these
objects (TVLM513-46546, 2MASSJ0036+18, LSRJ1835+3259,
2MASSJ1047539+21 \citealt{rw12, hallinan08}) emit periodic, 100\%
polarised radiation, normally detected at 4-8 GHz with spectral
luminosities of $\approx 10^{13.5}$~\ergshz \citep{had+06, rw12}.
These pulses are caused by the cyclotron maser instability (CMI), the
emission mechanism that occurs on Jupiter \citep{t06, nbc+12, mmk+12}.
CMI emission requires a relatively tenuous population of energetic
particles confined to a relatively strong magnetic field; in
particular, the cyclotron frequency, $f_\text{ce}=e B / m_e c$, must
be much greater than the plasma frequency $f_\text{pe}=\sqrt{4 \pi e^2
  n_e/ m_e}$. Whenever detailed observations are available, the free
energy in the plasma is seen to be provided by electrons moving along
the magnetic field lines, which can originate in
magnetospheric-ionospheric (M-I) shearing and possibly plasma
instabilities.  These observations 
 suggest that BDs can self-generate stable, $\sim$kG-strength
magnetic fields \citep{b06b}. The underlying assumption is, however,
that enough free charges are present to form a plasma in these
extremely cold atmospheres ($<$2000 K).

Although this mechanism for emission is quite well characterised, and
can account for the polarised flaring behaviour, two of these dwarfs
also produce quiescent, moderately polarised emission. This indicates
that a second mechanism such as synchrotron or gyrosynchrotron
emission \citep{berger02, osten06, ravi11} may be occurring, or that
the CMI emission is becoming depolarised in some way as it crosses the
dwarf's magnetosphere \citep{hallinan08}. It has been suggested that
some of the variability seen in these sources may be due to variation
in the local plasma conditions \citep{stark2013}, perhaps linked to
magnetic reconnection events \citep{schmidt07} or due to other
sporadic charged events in the plasma (\citealt{hell2011, hell2011b,
  hell2013, bai2013}).

 It is still unknown what features distinguish radio "active" from
 radio "inactive" dwarfs. The relationship may depend on mass,
 effective temperature, activity, magnetic field strength and rotation
 rate. All the known radio "active" dwarfs have a high v$\sin{\rm
   i}$-value, indicating short rotation periods ($\sim$3hrs)
 \citep{antonova13, williams13a}. This may indicate a link between
 rotation rate and emission, but could also indicate a dependence on
 the inclination angle, $i$, instead of the velocity, thus making
 detection of radio emission dependent on the line of sight and the
 beamed radiation emitted \citep{hallinan08}.

\subsection{Observed variability in Brown Dwarfs}\label{ss:obsvariab}

The number of campaigns that search for spectro-photometric variability
has increased since it was first suggested by \citet{tinney99}. Such
varia\-bility would indicate non-uniform cloud cover
(e.g. \citealt{ackerman01}) and weather-like features, such as those
seen on Jupiter.
The majority of the early searches concentrated on L dwarfs,
suggesting that variability could be due to holes in the clouds
\citep{gelino02, ackerman01}. Surveys suggest that between 40--70\% of
L dwarfs are variable \citep{bailer01,gelino02,clarke02}, although
most of these surveys involve small numbers of objects and the authors
vary on what is considered a detection. The majority of these studies
were performed in the $I$ band where the amplitude of variability is
at the 1--2\% level \citep{clarke02, clarke08} on timescales of tens of
minutes to weeks, and is in general not periodic. There has now been a
shift towards using the near-IR for variability studies
\citep{enoch03,clarke08, khandrika13, girardin13, buenzli14,
  radigan14, wilson14} where the frequency of variability is estimated
to be $\sim$10--40\%. So far, however, high amplitude periodic
variability ($>3$\%) has been limited to the L-T transition objects,
while lower amplitude variability is detected in both early L and late
T dwarfs.

\citet{heinze13} reported sinusoidal variability of the L3 dwarf DENIS-P J1058.7-1548 in the $J$ and [3.6] micron bands, but no variability in the [4.5] micron band. They suggested that the variability may again be due to inhomogenous cloud cover, where the thickest clouds have the lowest effective temperature, but there may also be an effect related to magnetic activity (starspots, chromospheric or  aurorae) suggested by a weak H$\alpha$ detection. This result is similar to the findings on the radio loud M8.5 dwarf TVLM513-46546 \citep{littlefair08}. TVLM513  shows $i'$ and $g'$ band variability in antiphase, initially suggesting the variability is due to patchy dust clouds coupled with the object's fast rotation. However, more recent results show that the $g'$ and $i'$ bands are no longer correlated. The $g'$ band variability has remained stable, but the $i'$ band lightcurve has evolved. However, the optical continuum variability is in phase with the H$\alpha$ flux, again suggesting some magnetic processes are also occurring \citep{metchev13}.

The first T dwarf to be confirmed as variable was the T2.5 dwarf SIMP
J013656.5\- +093347 \citep{artigau09} which was determined to be
variable at the 50 mmag level in the $J$ band.  A rotation period of
2.4 days was determined from the lightcurves. More interestingly, the
lightcurve evolves from night to night (Fig. 2 in \citet{artigau09}),
perhaps due to evolving features such as storms.  \citet{radigan12}
studied a similar object, the T1.5 dwarf 2MASS2139, and again found
evidence of an evolving light curve in the $J$, and $K_s$ wavebands,
which after extensive analysis they attributed to heterogeneous clouds
with regions of higher opacity. In general, these objects exhibit
variable lightcurves, modulated as the object rotates, which evolve on
a period of hours \citep{apai13}, days (e.g. \citealt {artigau09,
  radigan12}) or even years \citep{metchev13}.

While the majority of studied T dwarfs lie at the L-T dwarf transition
region, some late T dwarfs ($>$T5) have also been determined to be
variable \citep{clarke08}. Initially, these atmospheres were thought
to be relatively cloud-free, however recent work by \citet{morley12}
suggests that sulphide clouds may exist there (see
Sect.~\ref{ss:diffclmo}). \citet{buenzli12} studied one such object
and determined a phase offset between different wavelengths while
sinusoidal variability was present on the rotation prior of the
object.
This phase shift is directly linked to the pressure (and hence cloud
constituents) probed at each of the observed wavelengths
\citep{marley10, morley14b}. The lower pressure regions (high
altitude) cause the highest phase lag with respect to the highest
pressure layers (lowest altitude) and may be as large as half a
rotation period. The authors attribute this lag to a change in the
opacities (gas or cloud) without a change in the temperature profile,
a change in the temperature-pressure profile without a change in the
opacity or a combination of the two resulting in a "stacked-cell"
atmosphere as seen in Saturn
(e.g. \citealt{fletcher11}). \citet{robinson14} interpret the phase
lag in the frame of their 1D model as being due to thermal
fluctuations. However, their model is unable to reproduce the
variability on a timescale of $\sim$hours, as seen in the
observations, claiming that a 3D model will be required to explore the
dynamics fully.

An alternative to ground based observations is to move into space,
minimising differential refraction and atmospheric effects. One of the
largest space based variability surveys to date was performed by
\citet{buenzli14} who studied 22 brown dwarfs ranging from L5 to T6
with $HST$. Six of these objects are determined to be variable with
another 5 marked as potentially variable. This survey was not
sensitive to objects with long periods, but it still suggests that the
majority of brown dwarfs have patchy atmospheres, and that there is no
spectral type dependence on the fraction of dwarfs that are variable.

Perhaps one of the best studied variable objects is Luhman 16B.
Luhman 16AB is the third closest system to the sun at only 2pc
\citep{luhman13} being a L7.5 and T0.5 dwarf binary. \citet{gillon13}
reported variability on a 4.87 hour period with strong night to night
evolution which was attributed to a fast evolving atmosphere on the
cooler T0.5 dwarf. The L7.5 dwarf is not found to be
variable. \citet{burgasser14} performed spectral monitoring of the
system, modelling the T dwarf using a two-spot model and inferring a
cold covering fraction of $\sim$30-55\% varying by 15-30\% over a
rotation period. This resulted in a difference of $\sim$200-400~K
between the hot and cold regions.  \citet{burgasser14} interpreted the
variations in temperature as changes in the covering fraction of a
high cloud deck resulting in cloud holes which expose the deeper,
hotter cloud layers. They also suggested the rapidly evolving
atmosphere may produce winds as high as 1-3 kms$^{-1}$ which is
consistent with an advection timescale of 1-3 rotation periods.  A new
analysis of this system, was produced by \citet{crossfield14} who used
Doppler imaging techniques to produce a global surface map, sensitive
to a combination of CO equivalent width and surface brightness, of the
T dwarf Luhman 16B. The map shows a large, dark mid-latitude region,
a brighter area on the opposite hemisphere located near the pole and
mottling at equatorial latitudes (Fig. 2 in
\citealt{crossfield14}). The authors interpreted the map in one of two
ways. Either the darker areas represent thicker clouds, obscuring the
hotter, inner regions of the atmosphere, and the bright regions
correspond to cloud holes providing a view of this warmer interior, or
the map shows a combination of surface brightness and chemical
abundance variations.  They predict that the high latitude bright spot
could be similar to polar vortices seen in solar system giant planets,
in which case it should be seen in future mapping of this object.

Another class of brown dwarfs that shows photometric variability are
those in close ($<$10 hrs) detached binary systems with white dwarfs
(WD0137-349, WD+L6-L8, P=116 min: \citealt{maxted06, burleigh06};
GD1400, WD+L6, P=9.98hrs \citealt{farihi, dobbie, burleigh11};
WD0837+185, WD+T8, P=4.2hrs: \citealt{casewell12}; NLTT5306, WD+L4-L7,
P=101.88 min: \citealt{steele13}; CSS21055, WD+L, P=121.73 min:
\citealt{beuermann13}).  These systems are likely tidally locked and
as a result one side of the brown dwarf is continually heated by its
much hotter white dwarf companion. Three of the five known systems
show variability in optical wavelengths at the 1\% level. One of these
objects, WD0137-349 has been studied extensively  in the
near-infrared \citep{casewell13}. The white dwarf in this system is 10
times hotter than the brown dwarf, resulting in a difference in
brightness temperature of $\sim$500K between the day and night sides,
likely causing vigorous motion and circulation in the atmosphere
(e.g. \citealt{showman13}). While the substellar objects in these
systems are brown dwarfs, and not extrasolar planets, their
atmospheres behave in a similar way, both absorbing the (mainly)
ultraviolet emissions from their host, and also reflecting the
incident light. There are brown dwarfs known in close orbits with main
sequence stars that are also irradiated (e.g. WASP-30b:
\citealt{anderson11}, Kelt-1b: \citealt{siverd12}), but as their host
stars are much more luminous, and the brown dwarf atmosphere scale
height is too small to allow transmission spectroscopy, these are much
more challenging to observe.

\section{Theory of brown dwarf atmospheres}\label{s:theo}

 Most, if not all, observational findings reported in the
  previous sections result from or are influenced by atmospheric
  processes. The following section will therefore summarize the
  physics as we expect it to occur in ultra-cool atmospheres of brown
  dwarfs, including planets and M-dwarfs (Sect.~\ref{s:theo1}). 
  Special emphasis will be given to cloud formation processes and
  their modelling.  Section~\ref {ss:diffclmo} contains a summary of
  different approaches to treat cloud formation as part of model
  atmospheres.

\subsection{The brown dwarf atmosphere problem}\label{s:theo1}

\begin{figure}
\hspace*{-1cm}\includegraphics[width=1.2\textwidth]{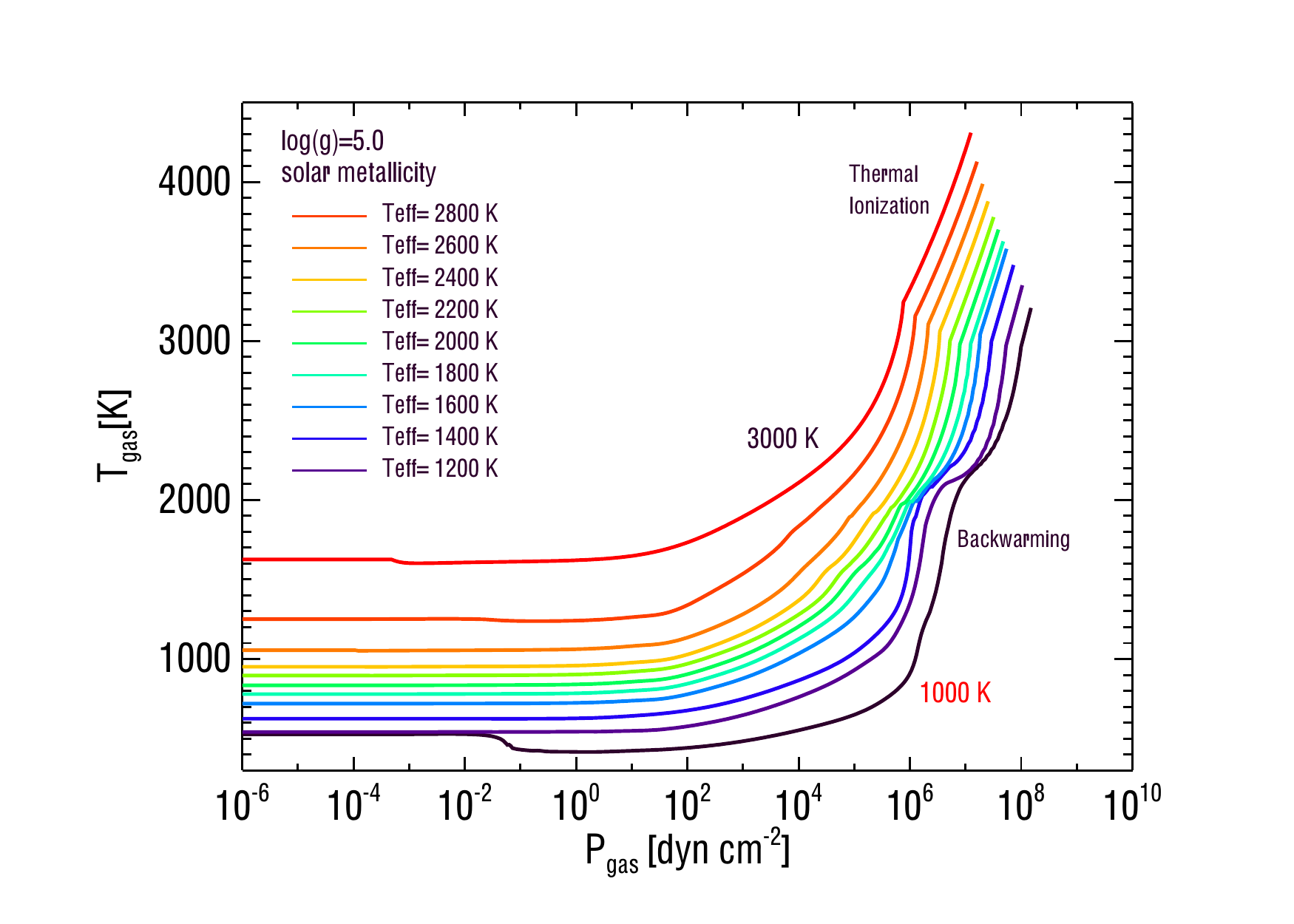}
\caption{The local temperature-pressure (T$_{\rm gas}$-p$_{\rm
    gas}$)-structure of brown dwarfs (log(g)=5.0) for effective
  temperature T$_{\rm eff}=1000\,\ldots\,3000$K. Shown are results
  from {\sc Drift-Phoenix} model atmosphere simulations for solar
  element abundances (\citealt{witte2009,witte2011}) [courtesy: Isabel
    Rodriguez-Barrera] }
\label{TgasPgas_DP}
\end{figure}

The atmosphere of a brown dwarf is composed of a cold gas with
temperatures $\approx 3000$K$\,\ldots\,<500$K, generally decreasing
outwards due to the upper atmosphere being an open boundary by
extending into space (Fig.~\ref{TgasPgas_DP}). Local temperature
inversions could occur due to locally heating processes: A locally
increased opacity could lead to radiative heating (e.g. backwarming,
see Fig.~\ref{TgasPgas_DP}) or cooling, Alfv{\'e}n wave propagation could
cause the occurrence of chromospheric structures, and irradiation
would provide an additional, external flux source.

 The description of the atmosphere of a brown dwarf requires to
 model the local thermodynamic (T$_{\rm gas}$ [K], p$_{\rm gas}$
 [dyn/cm$^2$]), hydrodynamic (v$_{\rm gas}$ [cm/s], $\rho_{\rm gas}$
 [g/cm$^3$]) and chemical properties (n$_{\rm x}$ [cm$^{-3}$], ${\rm
   x}$ - chemical species (ions, atoms, molecules, cloud particles))
 in order to predict observable quantities based on the radiative flux
 F$_{\lambda}$ [erg/s/cm$^2$/$\AA$]. The goal is to perform this task by
 using a minimum of global quantities that are observationally
 accessible like the resulting total radiative flux F$_{\rm tot}=\int
 F_{\lambda}\,d\lambda$ through the atmosphere. The classical 1D model
 atmosphere problem (e.g. \citealt{mihalas82}) is determined by the
 effective temperature T$_{\rm eff}$ (F$_{\rm tot}=\sigma T_{\rm
   eff}^4$, $\sigma$ [erg\,cm$^{-2}$s$^{-1}$K$^{-4}$] -
 Stefan-Bolzmann constant with the luminosity $L=4\pi R^2 \sigma
 T_{\rm eff}^4$ and the radius ($R$ [cm])), the surface gravity log(g)
 ($\log (g) = \log (GM/R^2)$,  $G$ [cm$^3$g$^{-1}$ s$^{-2}$] - gravitational
 constant, M - object mass) and the ele\-ment abundances of the
 object. Material quantities like the equation of state, and further
 chemistry and opacity data close this system of equations. Brown
 dwarf atmosphere models (\citealt{lunine1986, burrows1989, tsuji1996,
   tsuji2002, ackerman01, am2013, allard2001, burrows2002, gust2008,
   witte2009, witte2011}; for earlier references on M-dwarfs see
 \citealt{allard97}) that are widely applied to optical and IR
 observations (Sects.~\ref{ssOIR},~\ref{ss:obsvariab})
\begin{itemize}
\item[--] calculate the local gas temperature by solving the radiative and convective energy transport through a stratified medium in local thermal equilibrium (LTE) with an open upper boundary,
\item[--] assume flux conservation according to a total energy given by $T_{\rm eff}$\\ ($F_{\rm tot}=F_{\rm rad}+F_{\rm conv} = \sigma T_{\rm eff}^4$),
\item[--] calculate local gas-phase composition for a set of element
abundances to determine the opacity of the local gas (Mostly, chemical
equilibrium is assumed, in some simulations, CO/CH$_4$/CO$_2$ and  N$_2$/NH$_3$
are treated kinetically to better fit observed spectra.), 
\item[--] calculate the local gas pressure assuming hydrostatic equilibrium.
\end{itemize}

\noindent
Brown dwarfs  deviate from the classical model
atmosphere (e.g. textbook \citealt{mihalas82}) in
that clouds form inside their atmospheres.  During the formation
process, elements are consumed resulting in depleted gas-phase opacity
sources like TiO, SiO and others. The cloud particles are a
considerable opacity source which introduces a backwarming effect (see
Fig.~\ref{TgasPgas_DP}). Convection plays an important role in
transporting  material into regions that are cool enough
for the condensation processes to start or progress (see also Fig.~\ref{DustCircuit}). Convection is a
mixing mechanisms that can also drive the local gas-phase out of
chemical equilibrium if the transport is faster than the chemical
reaction towards the local equilibrium state (\citealt{noll1997}).

So far, the focus of brown dwarf model atmospheres was on one
dimensional simulations in the vertical ($z$) direction. This
assumption implies that the brown dwarf atmosphere is homogeneous in
the other two dimensions ($x, y$). Observation of irradiated brown
dwarfs (\citealt{casewell12}) suggest what is known for irradiated
planets since \cite{knutson2009}, namely, that global circulation may
also occur on brown dwarfs. We note, however, that the driving
mechanisms for global circulation are likely being dominated by
rotation in brown dwarfs compared to irradiation alone in close-in
planets (\citealt{zhang2014}).  Variability observations of
time-scales different than the rotational period
(Sect.~\ref{ss:obsvariab}) have long been interpreted as a sign for an
inhomogeneous cloud coverage of brown dwarfs.  {\it To summarize, the
  aim of building a model for an atmosphere is to understand the
  interaction between different processes and to calculate quantities
  that can be compared to experiments or observation.  The
  \underline{input quantities} are the global properties T$_{\rm
    eff}$, L$_{\star}$, log(g), mass or radius, element abundances,
  plus material constants for gas chemistry, cloud formation and
  opacity calculations. The \underline{output quantities} are details
  of the atmosphere like the local gas temperature, T$_{\rm gas}$ [K],
  the local gas pressure, p$_{\rm gas}$ [bar], the local convective
  velocity, {\sc v}$_{\rm conv}$ [cm/s], the local number densities of
  ions, atoms, molecules, $n_{\rm x}$ [cm$^{-3}$], local grain size,
  $a$ [cm], local number of cloud particles, $n_{\rm d}$ [cm$^{-3}$],
  local material composition of grains, the cloud extension, and many
  more details. Directly comparable to observations is the the
  resulting spectral surface flux $F_{\lambda}$.
  }

\subsection{The chemical repository of the atmosphere}

The chemical repository of an atmosphere, including atoms, molecules
and cloud particles, is determined by the element abundances available
throughout the atmosphere. Different wavelengths with different
optical depths probe different atmospheric layers with their specific
chemical composition. {\it Primordial element abundances}, that should
be characteristic for a young, hot brown dwarf, are determined by
where and when a brown dwarf formed as the interstellar element
abundances increase in heavy elements over time (\citealt{yuan2011})
and may depend on the star formation history of the brown dwarf's
birthplace (\citealt{henry1999,cheng2012}). The primordial abundances
should be preserved in the brown dwarf's interior and below any
atmospheric region that could be affected by cloud formation and
down-mixing of processed element abundances. Given the long life times
of brown dwarfs, we expect element sedimentation inside the brown
dwarf core similar to what we know from white dwarfs. The element
abundances that determine the spectral appearance of a brown dwarf are
{\it processed element abundances} and they differ from the primordial
values due to the effect of element depletion by cloud formation,
element enrichment by cloud evaporation and the convective mixing of
such chemically altered element abundances.

The primordial element abundances are almost always assumed to be the
solar element abundances or a scaling thereof. These values are
inspired by seismological measurements and 3D simulations to fit
high-resolution line profiles. The element abundance values determined
for the Sun depend a little on the method and/or simulation applied
(\citealt{pereira2013}, see also discussion in \citealt{hell2008}).
In principle, there is no reason why any star's element abundances
should precisely scale with the solar element abundances
(\citealt{berg2014}). \cite{allard97} summarized the chemical
composition in brown dwarf atmospheres as the basis of every model
atmosphere simulation.

\begin{figure}
{\ }\\*[-2.0cm]
\centering
\hspace*{-0cm}\includegraphics[width=\textwidth]{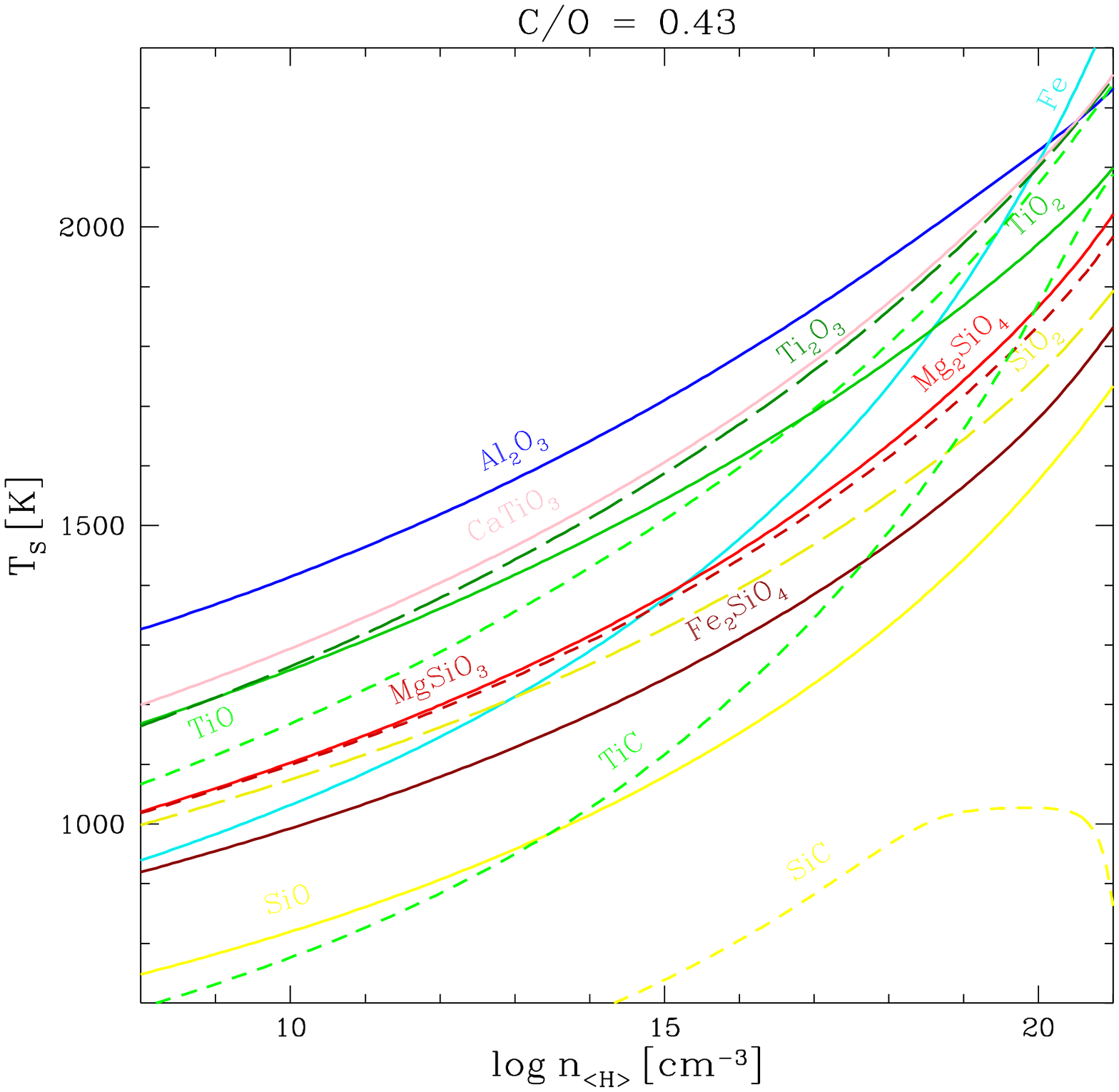}\\*[-4.6cm]
\hspace*{-0cm}\includegraphics[width=\textwidth]{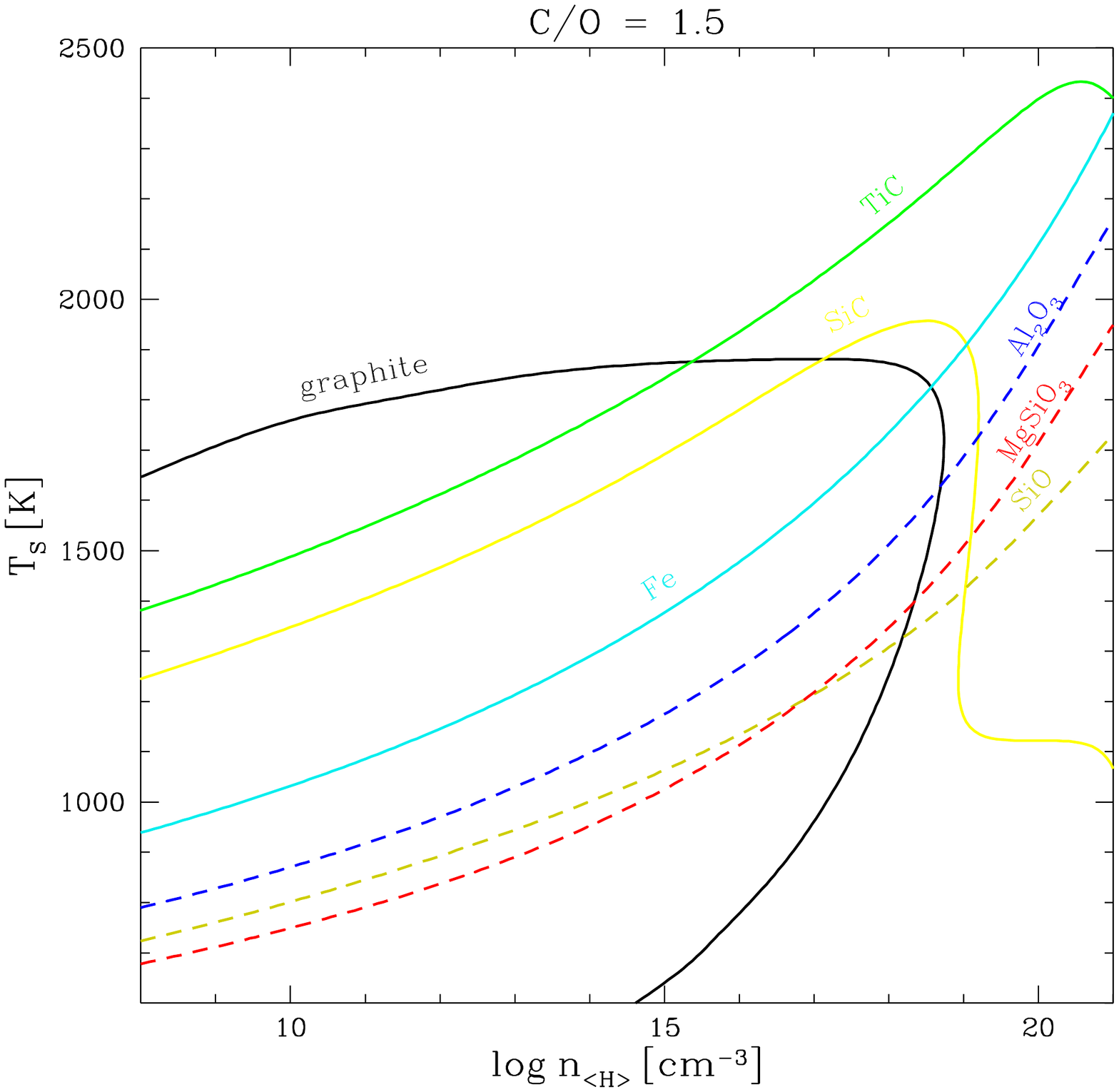}\\*[-2.5cm]
\caption{Thermal stability of various materials in an oxygen-rich
  (C/O$<1$; top) and a carbon-rich (C/O$>1$, middle) solar abundance
  gas. All curves show $(T_{\rm gas}, n_{<H>})$ where the
  supersaturation ratio $S_{\rm s}=1$ ($s$=TiO[s], Fe[s], graphite,
  $\ldots$; Eq.~\ref{eq:ss}). The materials will evaporate in the
  parameter space above each curve. In the case of graphite, the
  evaporation parameter space is above and to the right of of the
  curve [courtesy: P. Woitke].}
\label{Tsub}
\end{figure}

\subsection{Fundamental ideas on cloud formation }

The following section provides a summary of basic ideas of how clouds
form. The formation processes and underlying concepts are based on a
microscopic approach which, in the very end, will depend on our
quantum-mechanical understanding of chemical reactions leading to more
and more complex structures that describe the transition from the
gas-phase into the solid or liquid phase. Section~\ref{ss:diffclmo}
will summarize different approaches to cloud formation modelling that
are applied by different research groups to solve the brown dwarf
atmosphere problem.

\begin{figure}[ht]
\centering
\hspace*{-0cm}\includegraphics[width=\textwidth]{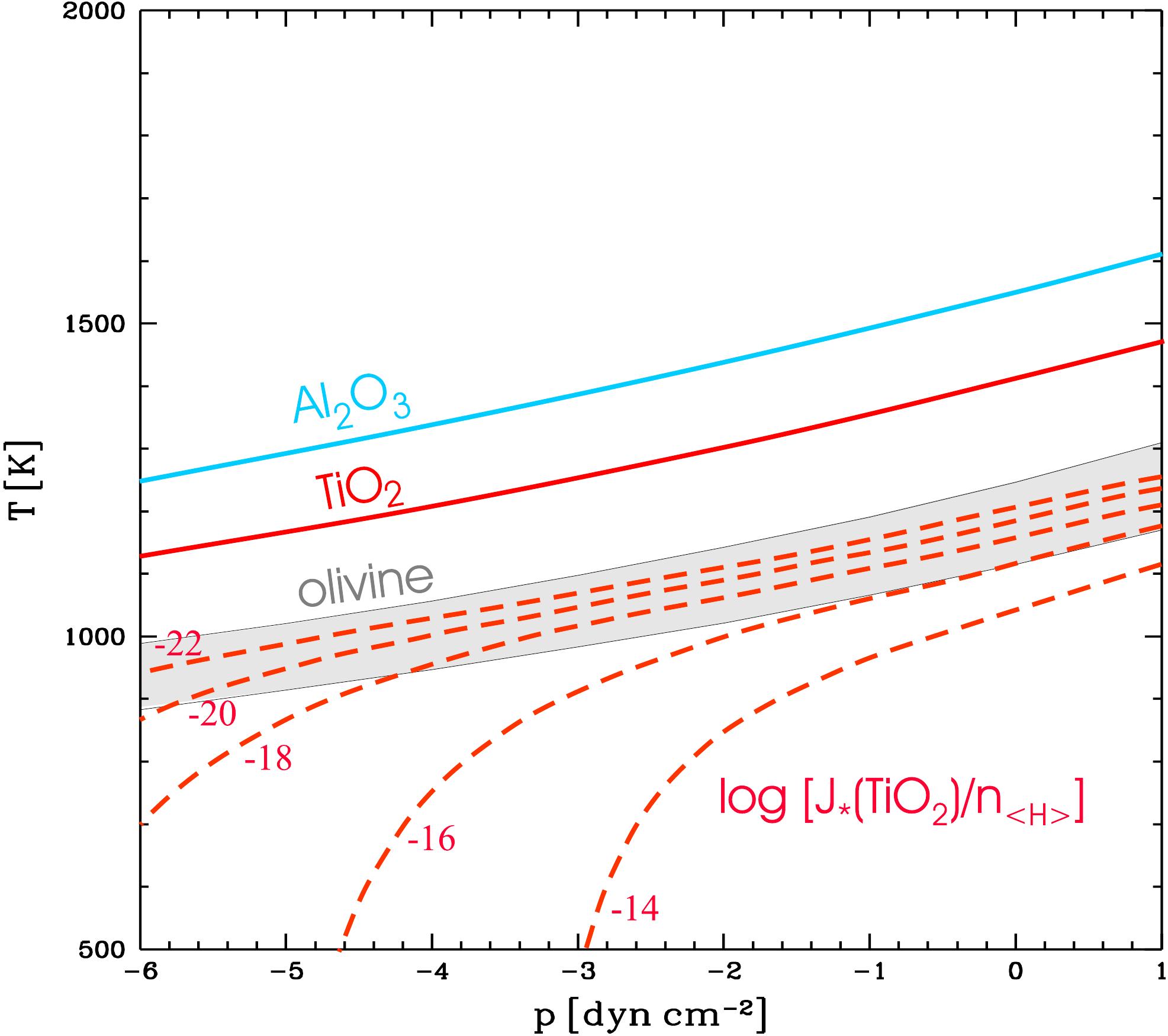}
\caption{A considerable supersaturation is required for the seed
  formation processes to occur. The seed formation rate of TiO$_2$
  (nucleation rate $J_*$, dashed red line) is highest far below the
  thermal stability curve for the solid material TiO$_2$[s] (solid red
  line).  Thermal stability for Al$_2$O$_3$[s] and the olivines in a
  solar abundance gas are shown for comparison. [courtesy: P. Woitke]}
\label{TsubJ*TiO2}
\end{figure}

\subsubsection{Thermal stability}
The concept of thermal stability is used in all but one cloud model
to determine if a cloud exists in an atmosphere. Figure~\ref{Tsub}
shows thermal stability curves usually used for this procedure, and
which are often called 'condensation curves' in the
literature. 
Such a
material is in {\it phase equilibrium} which is described by the
supersaturation ratio $S_{\rm s}=\,1$ of a material $s$ ($s\,$= SiO[s],
TiO$_2$[s], MgO[s], Fe[s], Mg$_2$SiO$_3$[s], $\ldots$; [s] referring to 'solid') with the
supersaturation ratio being defined as
\begin{equation}
\label{eq:ss}
S_{\rm s}=\frac{p_{\rm x}(T_{\rm gas}, p_{\rm gas})}{p_{\rm sat, s}(T_{\rm s})}.
\end{equation}
$p_{\rm x}(T_{\rm gas}, p_{\rm gas})$ is the partial pressure of the
growing gas species $x$, and $p_{\rm sat, s}(T_{\rm s}$) is the
saturation vapour pressure of the solid $s$. The application of the
law of mass action to $p_{\rm x}$ shows that $S_{\rm s}$ is
well-defined no matter whether the monomer of solid $s$ exists in the
gas-phase or not (\citealt{hewo2006}). Hence, the concept of thermal
stability does not allow to investigate if and how a particular
condensate does form.  Figures~\ref{Tsub} demonstrate below (i.e., $\leqq$) which temperatures  a material
would be thermally stable in an oxygen-rich (top) and a carbon-rich (bottom) environment.

\subsubsection{Cloud formation processes}\label{ss:clp}

\begin{figure}[ht]
\centering
\hspace*{-0cm}\includegraphics[width=\textwidth]{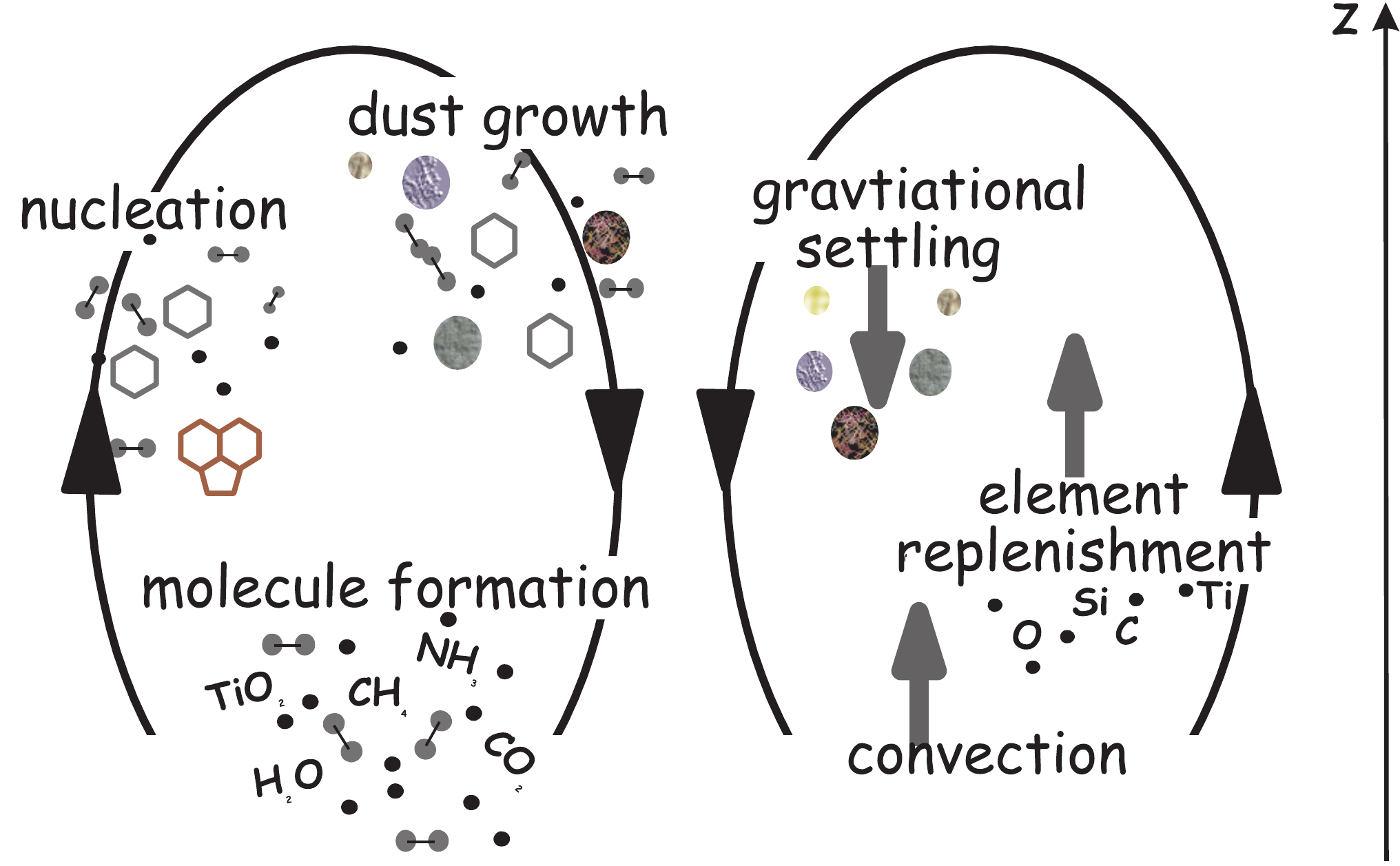}
\caption{The circuit of dust that determines cloud formation: cloud particle formation (nucleation, growth) $\rightarrow$ gravitational settling (drift)  $\rightarrow$ element depletion \& element replenishment by convective mixing (\citealt{woi2004}).}
\label{DustCircuit}
\end{figure}

Cloud formation (Fig.~\ref{DustCircuit})
starts with the formation of condensation seeds in
brown dwarfs and giant gas planets where no tectonic processes can
provide an influx of dust particles into the atmosphere.\footnote{The
  formation of weather clouds on Earth involve water condensation of
  pre-existing seeds particles ({\it condensation nuclei}) which
  origin from volcano outbreaks, wood fires, ocean salt spray, sand
  storms, and also cosmic-ray-induced ion-ion cluster reactions (see
  CERN CLOUD experiment). Noctilucent clouds in the upper Earth
  atmosphere, however, require the recondensation of meteoritic
  material to understand their existence (\citealt{saunders2007}).}
This process is a sequence of chemical reactions through which larger
molecules form which then grow to clusters and eventually, a small,
solid particle emerges from the gas phase. Such reaction chains have
been extensively studied in soot chemistry pointing to the key role of
PAH (polycyclic aromatic hydrocarbons) in carbon-rich environments
(\citealt{goeres1993}). The modelling aspect of such chemical paths is
greatly hampered by cluster data being not always available for a
sensible number of reactions steps, and computational chemistry plays
an important role for our progress in astrophysics' cloud formation
(e.g. \citealt{catlow2010}). Figure~\ref{TsubJ*TiO2} demonstrates that
the seed formation process requires a considerable supersaturation of
the respective seed forming species: The seed formation rate, $J_*$,
peaks at a far lower temperature than the thermal stability of the
same material suggests. This is very similar to Earth where water
vapour condensation on ions requires a supersaturation of
400\%. \cite{iraci2010}, for example, demonstrate that equilibrium
water ice formation is impossible on Mars.

Once condensation seeds have formed, other materials are already
thermally stable (Fig.~\ref{Tsub}) and highly supersaturated (Fig. 1
in \citealt{helling2008}). This causes the growth of a substantial
mantle via dust-gas surface reactions. The cloud particles forming in
an oxygen-rich gas will therefore be made of a mix of all available
materials as many materials become thermally stable in a rather narrow
temperature interval (top panel, Fig.~\ref{Tsub}). Once the cloud
particles have formed, other intra-particle collision processes may
alter the particle size distribution. Such collisions depend on the
momentum transfer between particles and may result in a further growth
of the particle or in destruction (\citealt{guettler2010,
  wada2013}). Collisions between charged grains may lead to an
acceleration of the coagulation process in brown dwarf atmospheres
(e.g. \citealt{konopka2005}).

\begin{figure}
\centering
\includegraphics[width=1.0\textwidth]{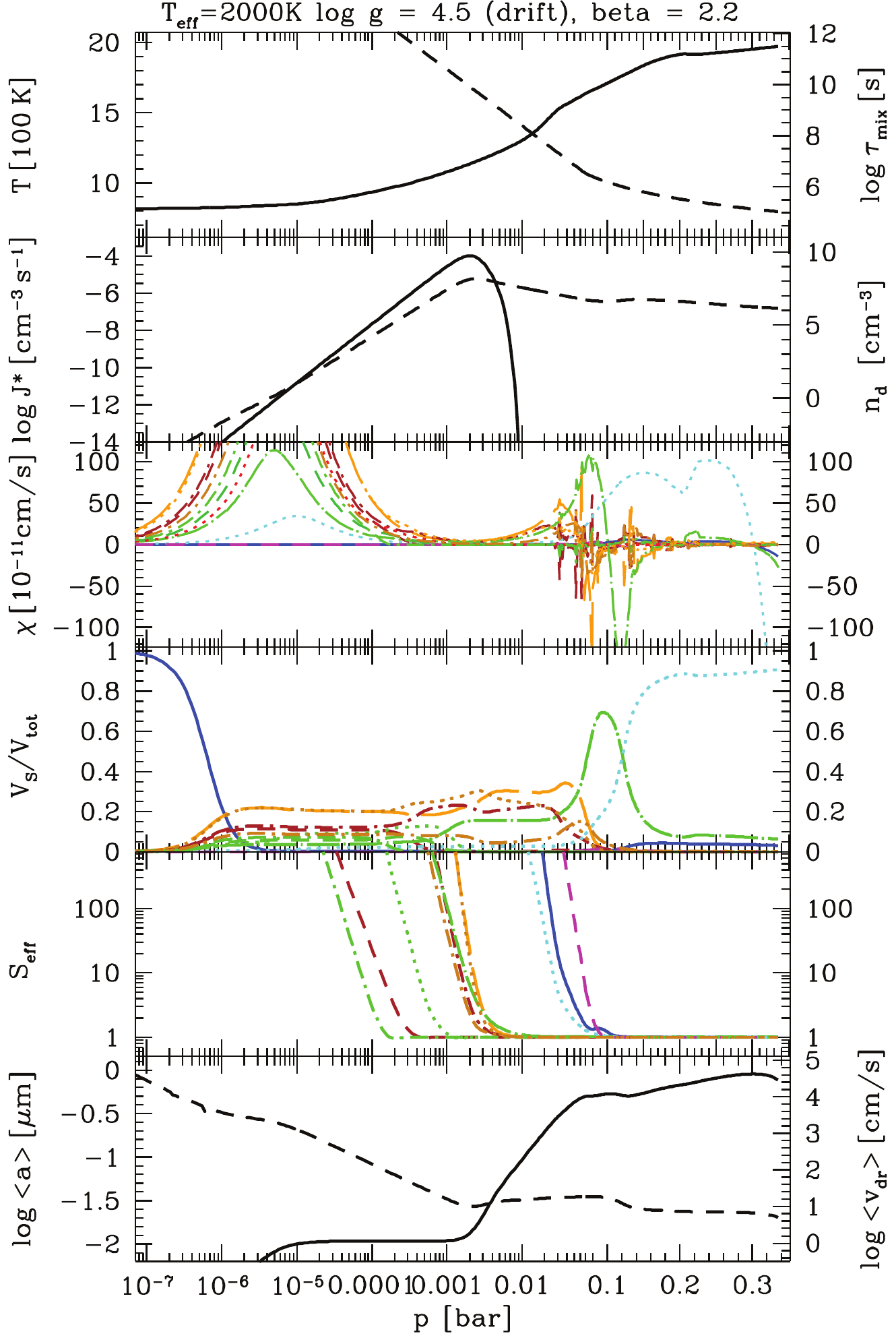}\\*[-0.2cm]
\caption{Cloud model results as part of a {\sc Drift-Phoenix}
  atmosphere model (T$_{\rm eff}=2000$K, $\log$(g)=4.5, solar
  metallicity). {\bf 1st panel:} local gas temperature $T_{\rm gas}$
  [K] (solid), convective mixing time scale $\tau_{\rm mix}$ [s]
  (dashed); {\bf 2nd panel:} seed formation rate $J_*$
  [cm$^{-3}$s$^{-1}$] (solid), cloud particle number density $n_{\rm
    d}$ [cm$^{-3}$] (dashed); {\bf 3rd panel:} grain growth velocity
  $\chi_{\rm s}$ [cm s$^{-1}$]; {\bf 4th panel:} material volume
  fraction $V_{\rm s}/V_{\rm tot}$ [\%]; {\bf 5th panel:} effective
  supersaturation ratio $S_{\rm eff}$ for each material $s$; {\bf 6th
    panel:} mean grain size $\langle a \rangle$ [$\mu$m] (solid),
  drift velocity v$_{\rm dr}$ [cm s$^{-1}$] (dashed). The different
  colours refer to the same different solid in each of the
  panels. {\small The subscript $s$ refers to the different condensate
    materials: $s$=TiO$_2$[s] (blue), Mg$_2$SiO$_4$[s] (orange,
    long-dash), SiO[s] (brown, short-dash), SiO$_2$[s] (brown, dot -
    short-dash), Fe[s] (green, dot - long-dash), Al$_2$O$_3$[s] (cyan,
    dot), CaTiO$_3$[s] (magenta, dash), FeO[s] (green, dash), FeS[s]
    (green, dot), Fe$_2$O$_3$[s] (green, dot - short-dash), MgO[s]
    (dark orange, dot - short-dash), MgSiO$_3$[s] (dark orange, dot).}
}
\label{DP2004.5_solar_struc}
\end{figure} 

\subsubsection{Some results on cloud {\it formation}}\label{ss:cloudresults}

We use modelling results from Helling \& Woitke to demonstrate
the origin of basic cloud properties and feedback mechanisms that
determine the formation of clouds. Refined results of a {\sc
  Drift-Phoenix} atmosphere simulation (\citealt{witte2011}) are used for this purpose.

\begin{figure}[ht]
\centering
\includegraphics[width=0.7\textwidth]{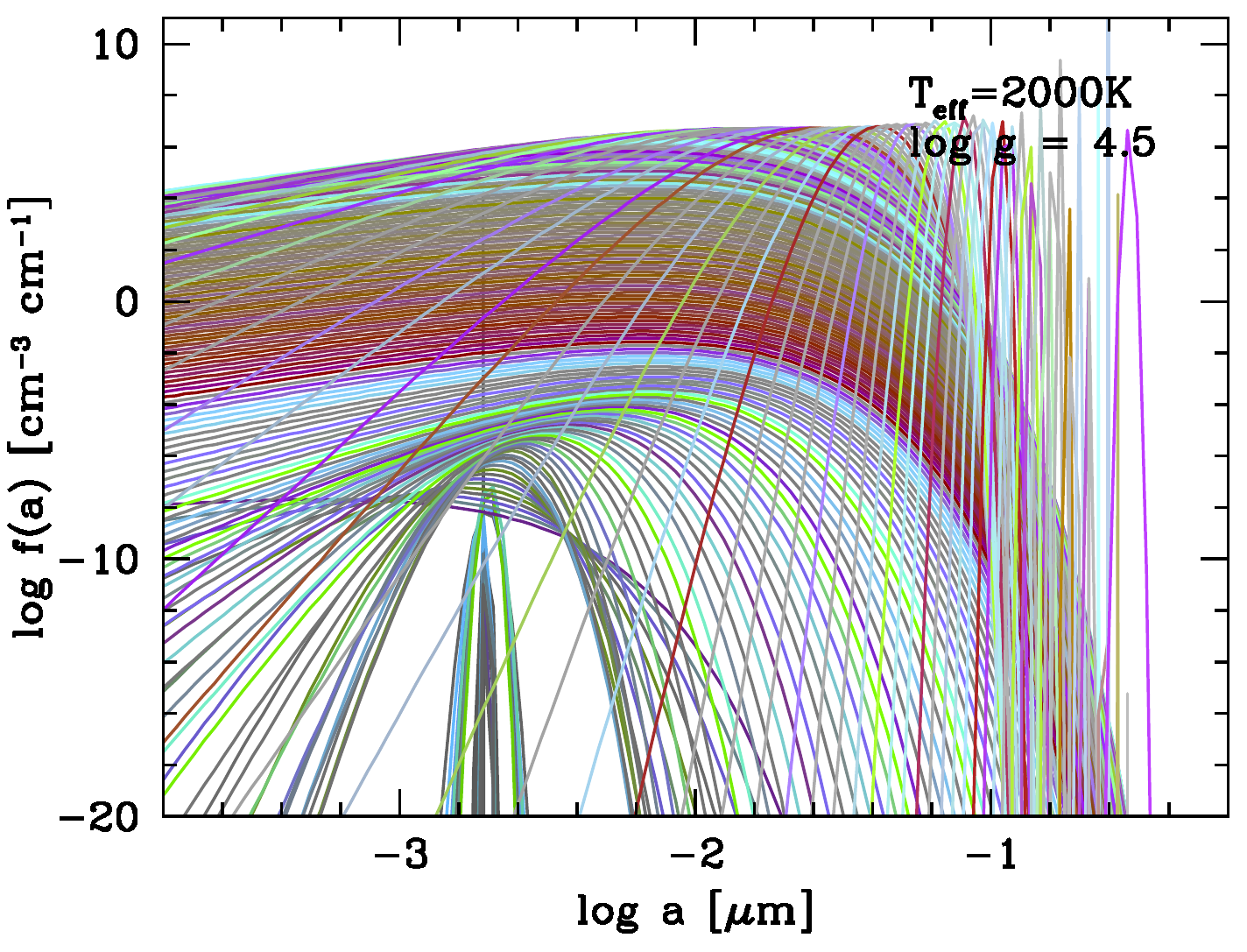}
\caption{Grain size distributions, $f(a)$ [cm$^{-3}$ m$^{-1}$], differ
  for different atmospheric layers due to different cloud formation
  processes contributing and/or dominating. The plot shows the
  evolution of the grain size distribution from the start of the cloud
  formation by seed formation (narrow gray-ish peaks), the continues
  production of seed particles which simultaneously grow (peaks
  growing in height and moving to the right towards larger grain
  sizes; blue colours), and the growth dominated distributions (peaks
  have constant heights and keep moving to the right; purple colours).
  $f(a)$ is shown for the same atmosphere model like in
  Fig.~\ref{DP2004.5_solar_struc}.}
\label{DP2004.5_solar_fvona}
\end{figure}

The upper boundary of the cloud is determined by the formation of
condensation seeds. Figure~\ref {DP2004.5_solar_struc} demonstrates
that a substantial grain growth can only set in when condensation
seeds have formed despite of extremely high supersaturation of
potentially condensing materials (5th panel).  The seed formation
rate, $J_*$ [cm$^{-3}$s$^{-1}$] (2nd panel, solid), is calculated for
TiO$_2$-nucleation for which cluster data exists (see 
\citealt{helfom2013}). It determines furthermore the number density of
cloud particles, $n_{\rm d}$ [cm$^{-3}$] (2nd panel, dashed) in the
entire cloud. Below a shallow TiO$_2$[s] layer, all thermally stable
materials grow almost simultaneously (3rd and 4th panel) producing
core-mantle cloud particles with a mixed mantle composition.

A closer inspection of the material volume fractions ($V_{\rm
  s}/V_{\rm tot}$) and the grain growth velocity ($\chi_{\rm s}$) reveals
a changing material composition in the cloud with height. The reasons
are element consumption and thermal instability: The condensates can
not grow further if element depletion causes a sub-saturation, or
evaporation sets in if the local temperature is too high. The result
is that also the individual supersaturation ratios approach one, but
each at a different temperature (i.e. atmospheric height in 1D models).

Element depletion affects also the seed formation (mainly due to Ti). No more new seed particles can
form below a certain height as the increasing temperature hampers the clusters' thermal stability. All cloud particles that exist below this
height have rained in from above. These particles fall into a gas of
increasing density and temperature. The increasing gas density causes
an increasing mean grain size ($\langle a \rangle$) until the grains
fall faster than they can grow. A part of the lower cloud has
therefore an almost constant mean grain size. Below that, the
temperature is too high and even high-temperature condensates like
Fe[s] and Al$_2$O$_3$[s] evaporate. The thermal stability of the most
stable materials determines the cloud's lower boundary. The lower edge
of the cloud is made of large particles that consist of very heat-resistant
materials like Al$_2$O$_3$[s] with inclusions of Fe[s] and TiO$_2$[s].

\subsubsection{Why do we need a cloud model?}
Important input quantities for model atmosphere simulations, and for
retrieval me\-thods, are opacity data for the gas-phase and for the
cloud particles. Molecular line lists have been a big issue for a long
time (\citealt{allard97, hill2013}). But the calculation of the gas
phase absorption coefficient requires also the knowledge of the number
density of the absorbing species which is determined by the element
abundances of the constituting elements. The element abundances are
strongly influenced by how many cloud particles form of which
composition and where in the atmosphere.  A detailed cloud model is
therefore needed to calculate how many cloud particles deplete the gas
where and of which elements. \cite{helling2008} detail in their Fig. 7
the changing [Ti/H], [Si/H], [Fe/H], etc abundances with atmospheric
height and global parameters when cloud formation is considered.

The cloud opacity is determined by the size distributions of the cloud
particles and their material composition as well as the optical
constants (refractory index).  The cloud opacity changes with height
because of the height dependent grain size distribution, but also the
material composition of the cloud particles changes with height
(Figs.~\ref{DP2004.5_solar_struc},~\ref{DP2004.5_solar_fvona}).
Figure~\ref{DP2004.5_solar_fvona} depicts the number of cloud
particles for each atmospheric layer considered in the underlying
atmosphere model: each atmospheric layer is characterized by one
curve. The distribution of the number of particles (denoted by $f(a)$)
also changes with atmospheric height. The evolution of the cloud
particle size distributions is determined by the cloud formation
processes summarized in Sect.~\ref{ss:cloudresults}. The delta-like
distributions (dark green curves) are characteristic for the top of
the cloud where the nucleation process dominates and surface growth is
not yet efficient (compare also panel 2 \& 3 in
Fig.~\ref{DP2004.5_solar_struc}). The distributions, $f(a)$, broaden
when grain growth becomes efficient and $f(a)$ increases in height when
nucleation takes place simultaneously (wide purple curves). This is
the case just below the cloud top. Once the size distributions 'move'
in grain-size space towards the right (towards larger grain sizes)
with a constant peak value, the nucleation has stopped. This is
indicative for those cloud regions where the cloud particles rain into
deeper atmospheric layers. The deeper cloud layers are characterized
by narrow size distributions of large grains as all cloud particles in
that layer had time to grow. Eventually, the distribution functions
move back into the small-grain region of
Fig.~\ref{DP2004.5_solar_fvona} because the cloud particles evaporate.

\begin{figure}
\centering
\hspace*{-0.4cm}\includegraphics[width=0.75\textwidth]{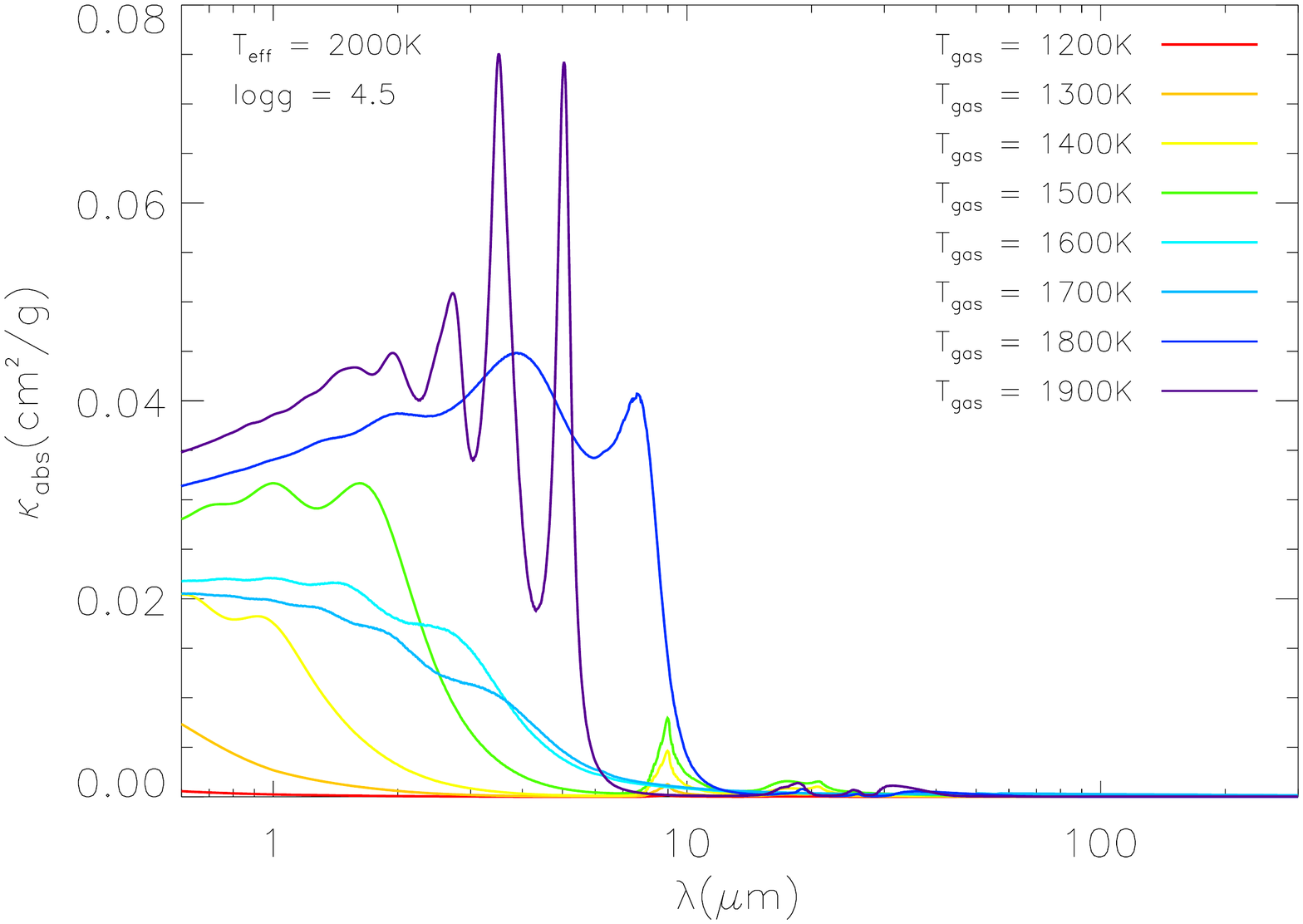}\\*[-1.3cm]
\hspace*{-0.4cm}\includegraphics[width=0.75\textwidth]{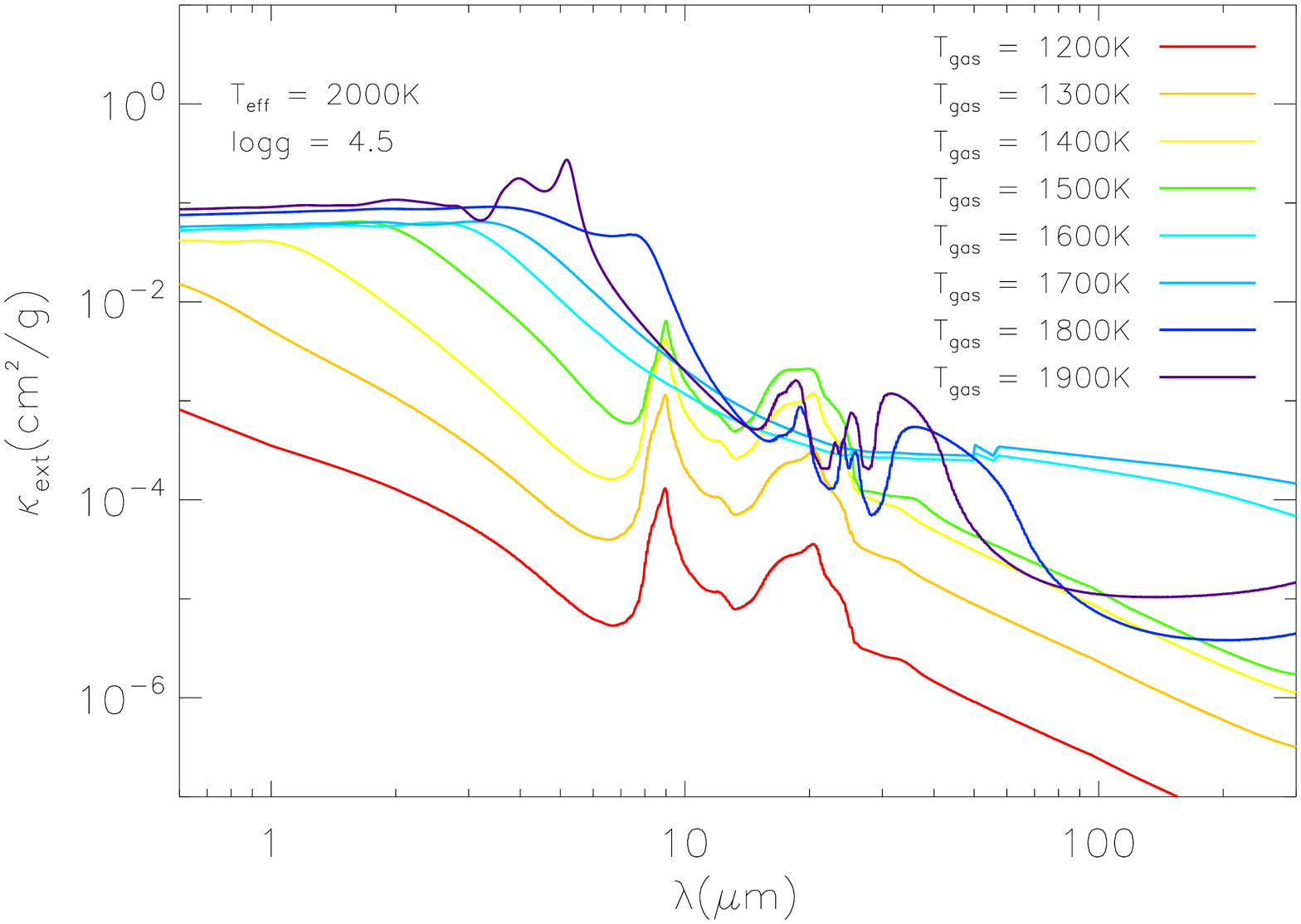}\\*[-0.7cm]
\includegraphics[width=0.61\textwidth]{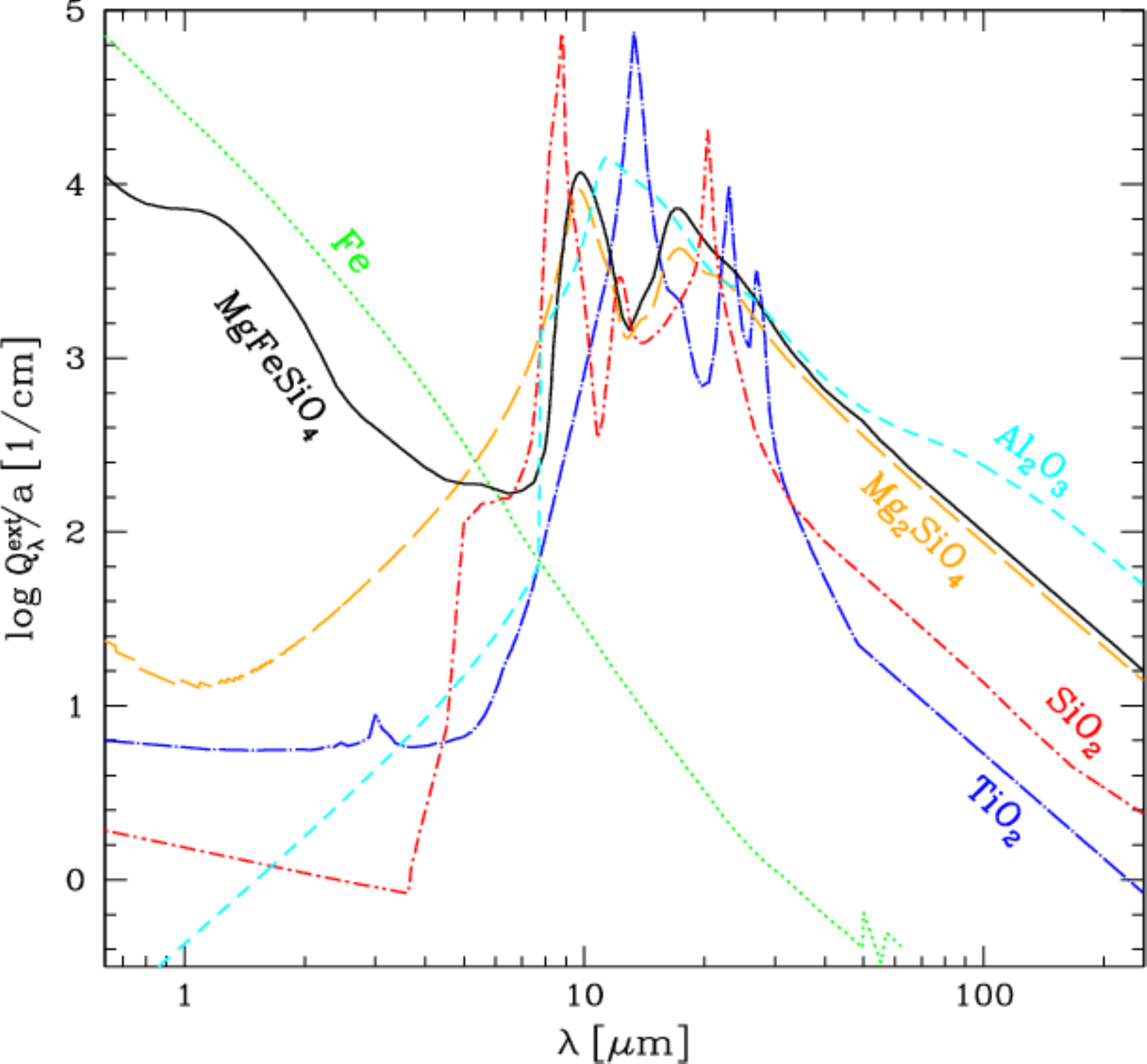}
\caption{{\bf Top two:} Cloud opacity, $\kappa^{\rm dust}_{\rm abs}(\lambda)$ and $\kappa^{\rm dust}_{\rm ext}(\lambda)= \kappa^{\rm dust}_{\rm abs+scat}(\lambda)$ [cm$^2$
    g$^{-1}$], as function of wavelength, $\lambda$ [$\mu$m], for
  different layers for a warm brown dwarf (T$_{\rm eff}=2000$K,
  log(g)=4.5). Different colours indicate different layers in the {\sc
    Drift-Phoenix} model atmosphere with each layer having different grain
  sizes and material compositions (Figs.~\ref{DP2004.5_solar_struc},\ref{DP2004.5_solar_fvona}) [courtesy: Diana Juncher]. {\bf  Bottom:} Extinction efficiency as function wavelength in
  the small-particle limit of the Mie-theory (\citealt{woitke2006}).}
\label{DustOpacity_Teff2000}
\end{figure}

Figure~\ref{DustOpacity_Teff2000}
shows wavelength-dependent cloud opacities for individual cloud
layers.  The silicate absorption features appears clearly in the
low-temperature part of the cloud (compare lowest panel
Fig.~\ref{DustOpacity_Teff2000}). Scattering dominates the cloud
extinction shortward of $4\,\ldots\,9\mu$m depending on the cloud
particle sizes.  Hence, for both, the gas opacity and the cloud
opacity, a rather detailed cloud model needs to be applied to
determine the cloud particle sizes, their size distribution and their
material composition depending on the local thermodynamic properties
inside the atmosphere.

\subsection{Different approaches to describe cloud formation in atmosphere simulations}\label{ss:diffclmo}

Cloud models are an integral part of each brown dwarf (and planetary)
model atmosphere simulation as they determine the remaining element
abundances that define the local gas-phase composition of the spectrum
forming atmosphere layers.  In the following, we summarize the
different cloud models that are to date applied and published in model
atmosphere simulations of brown dwarfs (and planets and M
dwarfs). This section is an update of \cite{hell2008} and it includes
all brown dwarf cloud models that are part of an atmosphere
simulation. \cite{ackerman01} discuss some of the older cloud models
(\citealt{lunine1986, rossow1978}).  A more planet-focused review is
provided in \citep{marley2013}.  \cite{lunine1986} and in their next
paper \cite{burrows1989} were the first to introduce a cloud opacity
(or 'particulate opacity sources') into their atmosphere model which
served as outer boundary for brown dwarf evolution
models. \cite{tsuji1996} suggested that dust needs to be taken into
account as opacity source for atmosphere models with T$_{\rm
  eff}<$2800K.

The following cloud models are very different from any cloud
parametrization used in the classical retrieval methods. Retrieval
methods only use a radiative transfer code that is iterated until an
externally given set of parameters (local properties like molecular
abundances and gas temperatures) allows to fit a set of observed
properties (\citealt{mad2009, ben2012, bar2013, lee2013}).  Different
approaches are used to decide with which quality the parameter set
fits the observation. \cite{lee2013} and \cite{line2014} assume
Gaussian error for single-best fit solutions, \cite{ben2012} derive a
full probability distribution and credibility regions for all
atmospheric parameters. Both methods require a prior to start the
best-fit procedure in a multi-parameter space to ensure that the
global minimum can be found.


\paragraph{i) Tsuji model:} 
\cite{tsuji2001} suggested that condensates in cool dwarf atmospheres
are present in the form of layers with strict inner and outer
boundaries. The inner boundary, associated with a certain temperature
denoted by $T_{\rm cond}$, is related to the thermodynamical stability
of the cloud particles in the surrounding gas. The upper boundary,
parametrized by $T_{\rm cr}$, is related to the assumption that the
cloud particles must remain extremely small, because if they would grow
too large then they would otherwise settle gravitationally. For $T_{\rm
  cond} > T > T_{\rm cr}$, the particles are assumed to be constantly
forming and evaporating, thereby circumventing the problem of the
gravitational settling (\citealt{tsuji2001, tsuji2002, tsuji2004,
  tsuji2005}). \cite{yam2010, tsuji2011, sora2013} have applied their
{\sc Unified Cloudy Model} to the unique AKARI data set for brown
dwarfs.  \cite{sorahana} interpreted the mismatch with these models as
signature of chromospheric activity on brown dwarfs (see Sect.~\ref{ss:halfa}).

\paragraph{ii) Burrows et al. model:} 
\cite{cooper2003} (also \citealt{burrows2006}) assume chemical and
phase equilibrium to determine whether cloud particles of a certain
kind are thermodynamically stable in a solar composition gas. If
$p_{\rm x}(T_{\rm gas}, p_{\rm gas})=p_{\rm sat, s}(T_{\rm gas})$,
i.e. the material $s$ is thermally stable, the mean size of the
particle of a certain homogeneous composition $s$ is deduced from
local time-scale arguments (\citealt{rossow1978}), considering growth,
coagulation (also named coalescence), precipitation and convective
mixing.  $p_{\rm x}(T_{\rm gas}, p_{\rm gas})=p_{\rm sat, s}(T_{\rm
  gas})$ also determines the altitude of cloud layers of composition
$s$. The amount of dust is prescribed by a free parameter $S_{\rm max}
\approx 1.01$ (maximum supersaturation) which is the same for all
materials.  Thereby, the supersaturation ratio of the gases is fixed
throughout the atmosphere and the mass of cloud particles present in
the atmosphere scales with the saturation vapour pressure $p_{\rm sat,
  s}(T)$, which decreases exponentially with decreasing $T_{\rm
  gas}$. Consequently, the vertical cloud structure is a dust layer
with a strict lower boundary and an exponentially decreasing
dust-to-gas ratio above the cloud base.  \cite{burrows2011} use this
phase-equilibrium approach to search where $S_{\rm s}$=1.0 for
individual materials $s$. A cloud density function is distributed
around this local pressure parametrizing the geometrical cloud
extension. For each condensate $s$ the vertical particle distribution
is approximated by a combination of a cloud shape function and
exponential fall-offs at the high- and low-pressure ends.  The model
has been used, for example, in \cite{apai2013} to suggest the presence
of an upper, warm thick cloud and a lower, cool thin cloud as reason
for the observed atmospheric variability of two early L/T-transition
dwarfs (2M2139, SIMP0136; see Sect.\ref{ss:obsvariab}). Different
cloud shape functions are tested with constant vertical distribution
of particles above the cloud base (B-clouds, same as DUSTY models in
Allard et al. 2001), and the modal grain sizes per material are
adjusted to provided the best spectral fit.

\paragraph{iii) Marley et al. model:}
\cite{ackerman01} parametrize the efficiency of sedimentation of cloud
particles relative to turbulent mixing through a scaling factor,
$f_{\rm sed}$. Large values of $f_{\rm sed}$ describe rapid particle
growth and large mean particle sizes. In this case, sedimentation is
efficient, which leads to geometrically and optically thin
clouds. When $f_{\rm sed}$ is small, particles are assumed to grow
more slowly and the amount of condensed matter in the atmosphere is
larger and clouds are geometrically more extended.  Marley et
al. solve a diffusion equation that aims to balance the advection and
diffusion of each species vapor, $p_{\rm x}(T_{\rm gas}, p_{\rm
  gas})$, and condensate, $p_{\rm sat, s}(T_{\rm gas})$, at each layer
of the atmosphere. It balances the upward transport of vapor and
condensate by turbulent mixing with the downward transport of
condensate by sedimentation. The downward transport of each condensate
is parametrized by $f_{\rm sed}$ and the turbulent mixing by an eddy
diffusion coefficient, $K_{\rm zz}$ [cm$^2$ s$^{-1}$].  The partial
pressure of each condensate species, $p_{\rm x}$, is compared with the
condensate vapour pressure, $p_{\rm sat, s}$, the crossing of which
defines the lower boundary of the atmosphere above which this
particular condensate is thermally stable. The model, as all other aforementioned models,  assumes that each
material can form by homogeneous condensation. \cite{ackerman01}
compute a single, broad log-normal particle size distribution for an
assumed modal size that is intended to capture the likely existence of
a double-peaked size distribution. \cite{fort2008} apply this cloud
model to planetary atmosphere simulations and, for example, suggest
two classes of irradiated planets.  \cite{morley12} applied the Marley
et al. model to suggest the presence of an additional cloud layer by moving
higher into the atmosphere towards lower temperatures.

\paragraph{iv) Allard et al. model:}
Phase equilibrium between cloud particles and gas is also assumed in
this model (\citealt{allard2001}). This assumption, i.e.  $p_{\rm
  x}(T_{\rm gas}, p_{\rm gas})=p_{\rm sat, s}(T_{\rm gas})$, is used to determine the cloud base
for each condensate individually. Using the time scale for
condensation, sedimentation and coalescence
(\citealt{rossow1978}) in comparison to a prescribed mixing time-scale
allows to determine a local mean grain size for a given grain size
distribution (\citealt{allard2013}). The mixing-time scale described
the convective overshooting based on a mass exchange frequency guided
by \cite{ludwig2002}.  The {\sc BT-Settl} models were applied to learn
about variability in brown dwarf atmospheres, for example on Luhman 16
by \cite{crossfield14}.

\begin{table}
\label{tab:cloudmodels}
{\small 
\hspace*{-0.5cm}
\begin{tabular}{l|l|l|l|ll}
  &    grain  & grain  & gas &   \multicolumn{2}{l}{fitted }\\
  &    size  & composition & saturation &   \multicolumn{2}{l}{ parameters}\\
\hline
\multicolumn{5}{l}{{\bf \underline{Model atmosphere simulations}}}\\[0.2cm]
Tsuji$(^1)$         & $a=10^{-2}\mu$m & homog. & $S=1$ & {\it UCM}   & dust between\\
                       &                 &             &      & & $T_{\rm cr}<T<T_{\rm cond}$ \\[0.1cm]
 Allard \& Homeier $(^2)$ & $f(a)=a^{-3.5}$    & homog. & $S=1$ &{\it dusty} & full dusty model\\
               &          &             &       &{\it cond}    & dust cleared model \\
               &time scales dep.   & homog. & $S=1.001$ &{\it settl} & time scales\\
                                                              &                                 &              &             & $K_{\rm zz}$ & mixing for non-\\
                                                              &                                 &              &             &                      & equilibrium molecules\\[0.1cm]
Cooper, Burrows et al.$(^3)$ &  $f(a)\sim\big(\frac{a}{a_0}\big)^6$ & homog. &$S=1.001$ &  & dust between \\
 & $\times\exp\big[-6\big(\frac{a}{a_0}\big)\big] $ & & & & $P^{\rm cloud}_{\rm upper}$, $P^{\rm cloud}_{\rm lower}$ \\ [0.1cm]
 Barman $(^4)$ & log-norm. \! $f(a, a_0)$ & homog. & $S=1$ & $P_{\rm min}$, $a_0$ \\
Ackerman \& Marley $(^5)$   & log-norm. \! $f(a,z)$ & homog. & $S=1$ & $f_{\rm sed}$ & sedimentation \\
                                               &                                 &              &             & $K_{\rm zz}$ & mixing for non-\\
                                               &                                 &              &             &                      & equilibrium molecules\\
Helling \& Woitke$(^6)$   & $f(a, z)$           & mixed  & $S=S(z,s)$&&\\[0.3cm]
\multicolumn{5}{l}{{\bf \underline{Retrieval method (radiative transfer only + fit quality assessment)}}}\\[0.2cm]
Barstow, \!Irwin, Fletcher$(^7)$  & $a_1$ & homog. & -- &  \multicolumn{2}{l}{n$_{\rm mix}$(H$_2$O, CO$_2$, CH$_4$) }\\
+ Benneke \& Seager$(^8)$   &        log-norm. \!$f(a_2,z)$         &              &        &   \multicolumn{2}{l}{$\tau^{\rm cloud}(a_1, a_2)$, R$_{\rm Pl}$(@ 10 bar)}\\[0.1cm]
Lee, Heng \& Irwin$(^9)$  &  $a=$ const                & homog. & -- &    \multicolumn{2}{l}{$P^{\rm cloud}_{\rm up}$, $P^{\rm cloud}_{\rm down}$, $\tau^{\rm cloud}(Q_{\rm ext}(a))$}\\
+ Line, Fortney,    &   & & & \multicolumn{2}{l}{\small $\rightarrow$ more parameter possible}\\
Marley, \& Sorahana $(^{10})$          &                                 &              &             &   \\
\end{tabular}
\caption{
$(^1)$: \cite{tsuji1996, tsuji2002, tsuji2004, tsuji2005}, 
$(^2)$:\cite{allard2001,bar2001, allard2013, rossow1978},
$(^3)$:\cite{cooper2003, burrows2006}, 
$(^4)$:\cite{barman11}, 
$(^5)$:\cite{ackerman01, morley12}, 
$(^6)$:\cite{woi2003, woi2004, hewo2006, helling2008, witte2009},
$(^7)$:\cite{bar2013, fle2009},
$(^8)$:\cite{lee2013},
$(^9)$:\cite{ben2012},
$(^{10})$:\cite{line2014}.
}}
\end{table}

\paragraph{v) Barman model:} Also \cite{barman11} assume phase-equilibrium and find the lower boundary of the cloud (cloud base) where the atmospheric ($T_{\rm gas}$, $p_{\rm gas}$)-profile intersects the thermal stability curve of a condensate ($p_{\rm x}(T_{\rm gas})=p_{\rm sat, s}(T_{\rm gas})$). The particle sizes follow a log-normal distribution with a prescribed modal size, $a_0$. The adjustable parameter $a_0$ can vary between 1 and 100$\mu$m and is the same for each atmospheric height. Prescribing the particle sizes allows  to determine the number of cloud particles (equilibrium dust concentration).  The cloud height and density above the cloud base is determined by a free parameter $P_{\rm min}$. The equilibrium dust concentration is assumed if  $p_{\rm gas} \ge P_{\rm min}$,  and decays exponentially for $p_{\rm gas}< P_{\rm min}$. If $P_{\rm min} > p_{\rm sat, s}(T_{\rm gas})$, then the maximum dusty-to-gas ratio is lowered relative to the 
equilibrium concentration.

\paragraph{vi) Woitke \& Helling model:} This model is different from all above models i)-v) as it  kinetically  describes seed formation and growth/evaporation coupled to gravitational settling, convective mixing and element depletion by conservation equations following Sect~\ref{ss:clp}. These intrinsically time-dependent processes (\citealt{hell2001, woi2003})  are treated in a stationary approximation  of conservation equations (\citealt{woi2004, hewo2006, hell2008}) to allow the coupling with a model atmosphere code ({\sc Drift-Phoenix}, \citealt{hell2008b, witte2009,  witte2011}). The convective mixing with overshooting is parametrized according to  a mass exchange frequency (\cite{ludwig2002}).  {\sc Drift-Phoenix} model atmospheres were applied to study early universe, metal-deficient brown dwarfs (\citealt{witte2009}), and they have recently been used to explore ionization and discharge processes in ultra-cool, cloud-forming  atmospheres (\citealt{hell2011, hell2011b, hell2013, rim2013, stark2013}).

\medskip
\noindent
{\bf Model approach summary:}\\ -- All phase-equilibrium models (i-v)
assume that each condensate can form by homogeneous condensation,
hence, it is assumed that the monomer exist in the gas phase. \\
-- All phase-equilibrium models (i-v) adjust their element abundances
according to by how much the monomer partial pressure exceeds the
vapour pressure $p_{\rm s}>p_{\rm sat, s}$. A prescribed size
distribution allows the calculation of  the cloud particle sizes. Note that
the two prescribed distribution functions (power law and log
normal) differ strongly in their dust mass distribution due to the
relative contribution of the different sizes.\\ 
-- All phase-equilibrium models
(i-v) are relatively easy to implement.\\ --
Table~\ref{tab:cloudmodels} provides a comparison of all cloud models,
including free parameters for each cloud model.

\begin{figure}[ht]
\includegraphics[width=\textwidth]{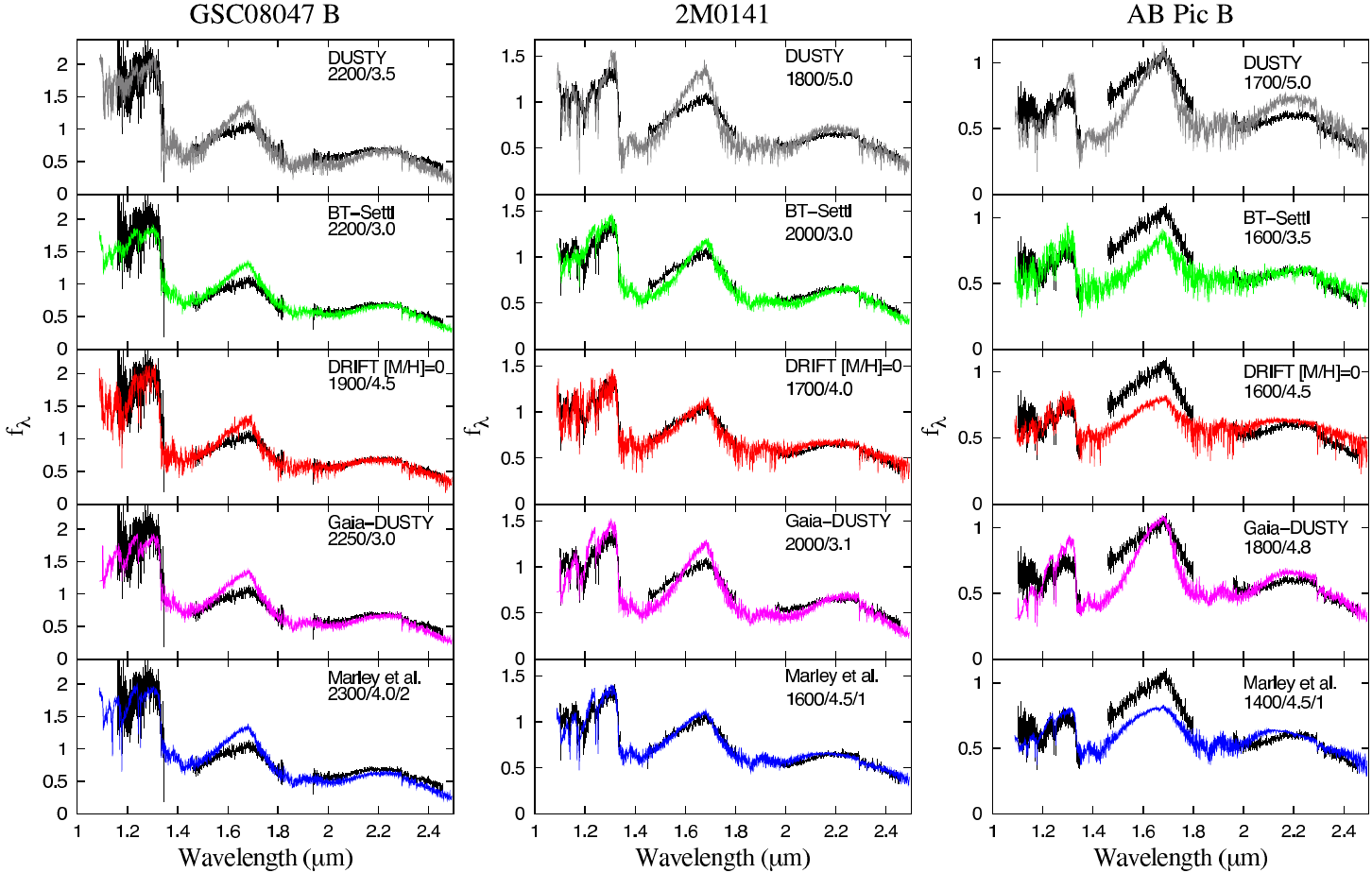}
\caption{\cite{pat2012} demonstrate spectral fits for varies Brown
  Dwarfs with synthetic spectra from different model
  families. Different model families were also used by \cite{dup2010} to
  provide a good error estimate on the derived global parameters. The figure shows that all model atmospheres found it challenging to fit brown dwarfs at the lowest T$_{\rm eff}$ depicted.}
\label{pat2012Fig8}
\end{figure} 

\section{Benchmarking atmosphere models and comparison to observations}\label{s:bench}
A 'benchmark' is per definition a well-defined test case which is
performed to understand differences in different approaches to the
same problem. Astronomers and Astrophysicists approach this differently.

The observational astronomy, which aims on benchmarking brown dwarf
models, follow the original idea that was inspired by stellar binary
systems for which one companion's parameters are well known
(\citealt{pin2006,burningham13}).  'Benchmark brown dwarfs' are
objects for which the distance (and thus luminosity), the age and the
metallicity can be determined from observations. Their radii and their
mass can be derived from evolutionary models if the age is well
constrained. The effective temperature and the surface gravity can now
be determined from the luminosity and the radius, the mass and radius,
respectively. With T$_{\rm eff}$, $\log$(g) and a known metallicity,
hence known element abundances, a model atmosphere is well constrained
and a synthetic spectrum can be produced. This model spectrum is now
compared to the binary target.  Obvious places to search for such
benchmark objects are open clusters as their age, composition and
distance can be well defined through observations of much more
luminous cluster members, and near-by kinematic moving groups with
well-defined membership, age and composition.

Theoretical astrophysics has two model systems to benchmark in our
context of brown dwarfs: the atmosphere models and the evolutionary
models. Both model systems are complex. The additional challenge is
that both systems are not independent because the model atmospheres
serve as outer boundaries during the run of evolutionary models.
Benchmarking both systems requires a substantial effort from different
research groups world-wide, and it is unlikely to happen in the near
future.
Ongoing areas of research, however, are dedicated comparison
studies that aim to demonstrate the difficulties in using model
atmosphere and/or evolutionary grids as 'black boxes'
(e.g. \citealt{sincl2010}), and to allow the observer community to use
an error estimate for model atmospheres (\citealt{boz2014}) and
evolutionary models (\citealt{southworth2009}). A sensitive handling
of such uncertainties is not only important for our immediate
understanding of brown dwarfs, but also for detecting planets around
brown dwarfs and M-dwarfs
(e.g. \citealt{rojas2013,triaud2013}.

Model atmosphere test are conducted in various ways. \cite{rajp2012}
test different model atmosphere families in finding the stellar
parameters for the late M-dwarf LHS 1070. They also including the
cloud-free MARCS models in their comparison with the cloud-modelling
{\sc Phoenix}-families (BTsettl, {\sc Drift}). \cite{dup2010} and
\cite{pat2012} compare the quality of fits to observations by model
atmospheres that describe the same physical problem, a cloud-forming
atmosphere (Fig.~\ref{pat2012Fig8}). The authors point out the
significant difference in the $H$-band spectrum for the different
models. This effects decreases with increasing T$_{\rm eff}$ but does
still impact the $J$-band for T$_{\rm eff}=1700$K. The primary
differences is in the cloud modelling. The cloud models differ in
details (grain sizes, material composition) that influence the local
opacity of the cloud and also of the gas phase through element
depletion. For example does a higher cloud opacity results in less
flux in the $J$-band as the water abundance decreases and, hence,
results in weakened water absorption bands. The approach presented in
\cite{dup2010} and \cite{pat2012} allows an error estimate of the
derived, global stellar parameters.  A more detailed comparison of
various brown dwarf model families is presented in
\cite{hell2008}. The focus of this paper is comparing cloud model
results for prescribed test cases. This also led to more
understanding about why the synthetic spectra from different model
families differ and which role the cloud modelling plays.

The Virtual Observatory pioneers the incorporation of grids of
different model atmosphere families into their data base. A comparison
between different model families can provide estimates of the expected
systematic errors, which is of interest for the outcome of space
missions.  \cite{sarro2013} performed the task of predicting how well
T$_{\rm eff}$ can be determined from GAIA data for Brown Dwarfs based
in the BTSettle model atmosphere grid.  The Virtual Observatory is now
capable of facilitating a multi-model family approach.

\section{Increasing completeness and increasing model complexity }\label{s:compl}

It may seem a big leap from a 1D atmosphere code to an atmosphere
model that allows the study of time-variable cloud cover, the formation of
photochemically driven hydro-carbonaceous macro-molecules, magnetic
interactions, and irradiation.  All these depend on global parameters like
rotational period, magnetic field strength, cosmic ray flux
etc. However, several aspects on this list start to emerge in
the literature.

\subsection{Multi-dimensional, dynamical atmosphere simulations}

First steps towards a multi-dimensional approach to cloud forming
brown dwarf atmospheres were made by 2D hydrodynamical simulations
that treat dust formation (nucleation, growth/evaporation, element
depletion), turbulence and radiative heating/cooling
(\citealt{hell2001, hell2004}), and by large-scale 2D
radiative-convection simulations that included the dust
growth/evaporation processes for a prescribed number of nucleation
seeds (\citealt{frey2010}). Both works suggest that clouds will not be
present as a homogeneous, carpet-like layer but that cloud particles
form, depending on the local temperature and density field,
intermittently resulting in patchy cloud structures.
\cite{robinson14} came forward with a similar suggestion of local
temperature variations to explain brown dwarf variability.
\cite{show2013} use a 3D approach to simulate a globally circulating
brown-dwarf atmosphere but excluding cloud opacities, turbulence, and
radiation. These authors suggest that a hydrodynamically induced
horizontal temperature variation of $\Delta T=50$K can lead to flux
variations of $\Delta F/F \approx 0.02-0.2$. Each of these models
addressed a different aspect of a multi-dimensional, dynamical
atmosphere simulations.  The challenges faced by all simulation is
illustrated by comparing the following time scales that are
characteristic for interacting processes. The below table demonstrates
that it is misleading to consider atmospheric processes as independent
from each other, but that for example cloud formation processes could
be re-ignited by transport processes like gravitational settling or
gas-mixing that provides new condensable material:
\begin{tabbing}
radiative cooling$^{\Diamond}$:\= \hspace*{0.8cm} \= $\tau_{\rm rad}(\rho_{\rm gas})$\,\hspace*{0.8cm}\=$=$\,\=$\big(\frac{4\pi \kappa}{c_{\rm V}}\frac{\partial B}{\partial T}\big)^{-1}$ \=\,\,\= =\,\,  0.5 $\ldots$ 100 days \\
\> \> \> \> \> \>{\small  (with dust \quad without dust)}\\
gravitational settling$^{\dag}$:\> \> $\tau_{\rm sink}(a, \rho_{\rm gas})$  \,\hspace*{0.5cm}\>$=$\,\>$\frac{H_{\rm p}}{{\rm v}_{\rm drift}}$ \>\,\,\= =\,\, 15 min $\ldots$ 8 month\\
\> \> \> \> \> \>\,\,  {\small ($a=100\mu$m$\ldots 0.1\mu$m)}\\
large-scale convection: \> \>$\tau_{\rm conv}(\nabla T_{\rm gas})$\>$=$\>$\frac{H_{\rm p}}{{\rm v}_{\rm conv}}$ \> \> =\,\, 20 min $\ldots$ 3.5h \\
diffusive eddy mixing$^{\ddag}$: \> \>$\tau_{\rm diff}$\>$=$\>$\frac{H^2_{\rm p}}{K_{\rm eddy}}$ \> \> =\,\, 3h$\,\ldots\,$ 3 yrs\\
grain growth: \> \> $\tau_{\rm gr}(T_{\rm gas}, \rho_{\rm gas})$\>$=$\>$\frac{a}{{\rm v}_{\rm gr}}$ \> \> =\,\, 0.1s $\ldots$ 1.5min\\
wave propagation: \> \> $\tau_{\rm wave}(T_{\rm gas}, \rho_{\rm gas})$\>$=$\>$\frac{H_{\rm p}}{u+c_{\rm s}}$ \>\,\,\= =\,\, 0.3s $\ldots$ 3s\\
seed formation: \> \> $\tau_{\rm nuc}(T_{\rm gas}, \rho_{\rm gas})$\>$=$\>$\frac{n_{\rm d}}{J_*}$\> \> $\approx$\,\,  $10^{-3}$s\\[0.2cm]
$^{\Diamond}$ {\small Please refer to Table 1 in \cite{hell2011} for definitions and values of the absorption coefficient $\kappa$, }\\
{\small   $c_{\rm V}$ specific heat capacity for constant volume and $B(T)$ the frequency integrated Planck function.}\\
$^{\dag}$ {\small Please refer to \cite{woi2003, woi2004} for definition of ${\rm v}_{\rm drift}$, ${\rm v}_{\rm gr}$, the mean }\\[0.0cm]
{\small grain size $a$, the number of cloud particles $n_{\rm d}$, and  the seed formation rate $J_*$. Typical}\\[-0.0cm]
{\small  values for all other quantities are applied, $H_{\rm p}\approx 10^6$cm.}\\
$^{\ddag}$ {\small $K_{\rm eddy}$ [cm$^2$s$^{-1}$] (or $K_{\rm zz}$) is the eddy diffusion coefficient ranging between $10^4\,\dots\,10^8$cm$^2$s$^{-1}$,}\\
{\small see \cite{bilg2013}}
\end{tabbing}

Diffusion, gravitational settling and convection have the longest time
scales in a brown dwarf atmosphere compared to chemical timescales for
cloud particle nucleation and growth. The wave propagation time scale
can be used as proxy for turbulence acting on small scales where
chemical processes would take place. The wave propagation takes still
$100\times$ longer, and hence, cloud particle formation would be given too
much time if a numerical scheme uses wave propagation to set
integration time-steps. This is problematic as the cloud particle
formation determines the remaining gas phase abundance which in turn
determines the local gas opacity and with that the local gas
temperature, and eventually, the energy transport and the spectral
flux. The radiative cooling time scale indicates the impact of the
radiative energy transport on the local hydrodynamics through the
energy equation. Depending on the local opacity, radiative cooling can
be very efficient, even causing local gas volumes to implode
(\citealt{hell2001}).

\subsection{Gas-phase non-equilibrium effects}

Deviations from local chemical gas-phase equilibrium in the upper
atmosphere are suggested to be caused by a rapid convective and/or
diffusive up-mixing of warm gases from deeper atmospheric layers
combined with a slow relaxation into chemical equilibrium
(\citealt{saum2000}).  Other processes that drive the local gas-phase
out of chemical equilibrium (and LTE) are photodissociation, or
ion-neutral chemistry initiated by cosmic ray impact (\citealt{rim2013}). The local
chemical composition of the atmosphere is derived from
extensive gas-phase rate network calculations under the influence of
vertical mixing and photodissociation. Most of these studies are
performed for irradiated planets (e.g. \citealt{moses2011,
  venot2012})\footnote{\cite{venot2012}'s chemical network is
  publicly available under http://kida.obs.u-bordeaux1.fr/.}.  A
height in the atmosphere (so-called quenching height) is derived above
which the gas kinetic reactions are too slow to considerably change
molecular abundances. This idea of modelling time-dependent gas
non-equilibrium effects has two shortcomings: First, the diffusive
eddy mixing coefficient $K_{\rm eddy}$ [cm$^2$s$^{-1}$] (or $K_{\rm
  zz}$) becomes an additional fitting parameter for model atmospheres
to observations (e.g. \citealt{cush2010}). Second, reaction rate
coefficients differ in the literature leading differences in
destruction time scale for example for C$_2$H$_2$ and C$_2$H$_6$
(\citealt{bilg2013}).  This is, however, a challenge faced by all
gas-kinetic approaches and considerable effort is ongoing to weed-out
the respective rates. \cite{rim2014} model cosmic ray transport
through an brown dwarf atmosphere and they demonstrate how galactic
cosmic rays influence the abundance of hydrocarbon molecules
through ion-neutral reactions.

\section{Closing the loop}

\begin{figure}[ht]
\centering
\includegraphics[width=1.0\textwidth]{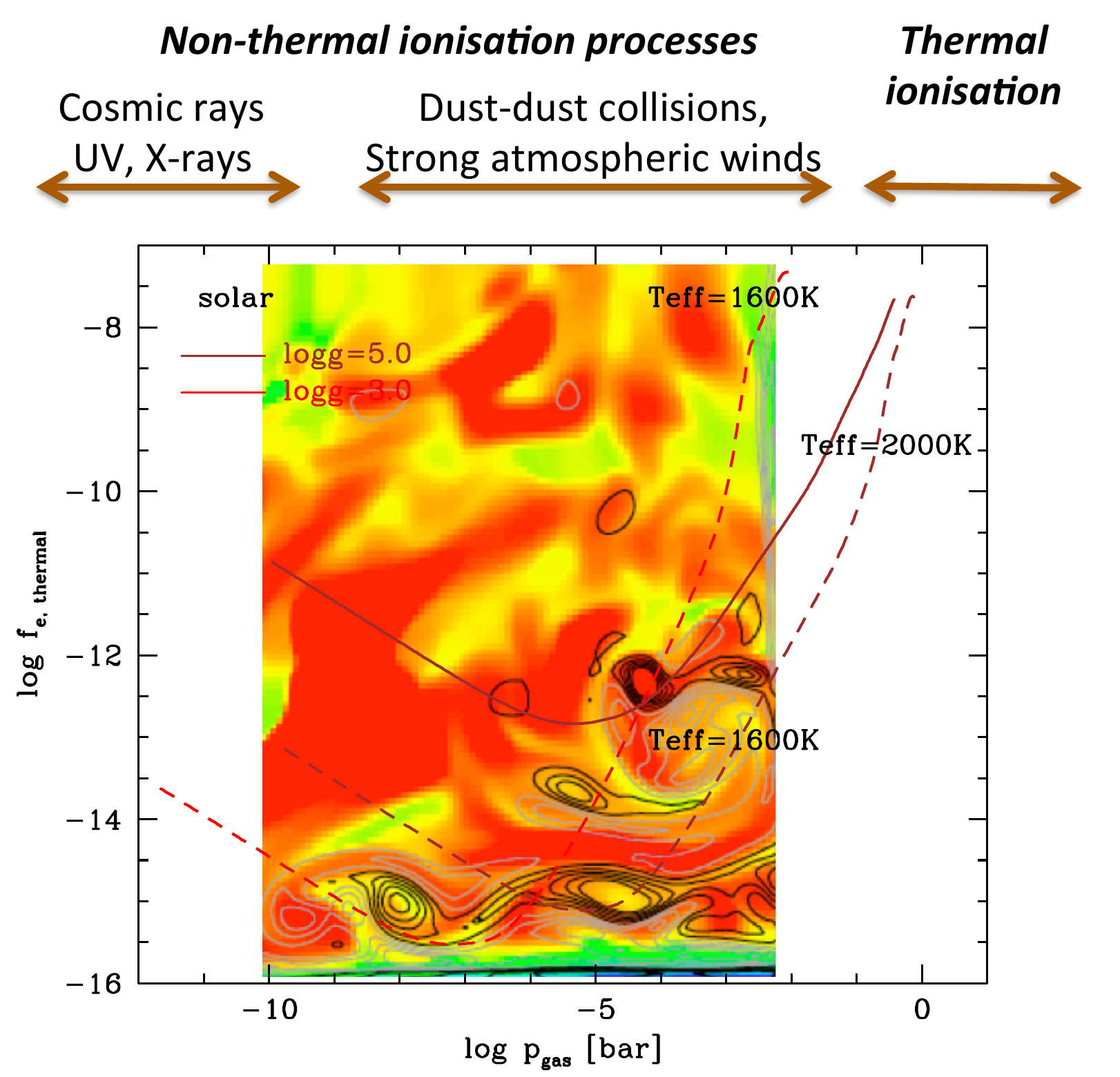}
\caption{A brown dwarf atmosphere is not only stratified by its cloud
  structure (2D colour overlay). Different non-thermal ionization
  processes occur in brown dwarf atmospheres (boxes: cosmic rays,
  dust-dust collisions, strong winds) which produce free charges
  through the atmosphere with varying efficiency. The local degree of
  thermal ionization (brown solid and dashed lines) is shown in the
  background for illustration, a 2D simulation of turbulent dust
  formation indicates where the cloud is located in the
  atmosphere. (The 2D colour-coded plot contains contour lines of the
  local vorticity; see Fig 8 in \cite{hell2004})
  }
\label{StratIon}
\end{figure}

The first L4 dwarf (GD 165B; \citealt{becklin88}) was discovered $\sim
30$ years ago. Since then, spectral classification of these very cool
objects led to the introduction of the new spectral classes L, T and
Y, with the Y dwarfs having T$_{\rm eff}$ typical for planets. Such
low temperature immediately suggest that brown dwarf atmospheres must
contain a chemically very rich gas from which clouds will form.  The
evolutionary transition from the L into the T dwarf spectral type are
associated with atmospheric variabilities which is attributed to
variable cloud coverage.  In parallel to the increasing number of
observations, atmosphere modellers adopted stellar atmosphere codes
for cooler gases by introducing cloud models and additional gas
opacity sources (e.g. CO$_2$, CH$_4$, NH$_3$). More complex processes
like kinetic gas chemistry, turbulence and multi-dimensional
hydrodynamic simulations were performed but with far less consistency
between the processes.\hspace{2cm} The picture that emerged for a
brown dwarf atmosphere is that of chemically very active gas that is
exposed to phase-changes, turbulence, high and low energy radiation.
It also provides a valuable path towards the understanding
for brown dwarfs as planetary host stars (\citealt{triaud2013}) and of
climate evolution on extrasolar planets.  More and more similarities
to planets arise: Radio and X-ray observations suggest brown dwarf
atmospheres to be ionized to a certain extend.  Theoretical studies on
ionization processes support this idea by demonstrating that clouds in
brown dwarfs will be charged (\citealt{hell2011b, hell2011}), that
clouds can discharge in form of lighting (\citealt{bai2013}), that
Cosmic Rays can ionize the upper atmosphere and the upper part of the
cloud (\citealt{rim2013}, and that hydrodynamic winds can provide a
source for gas-ionization (\citealt{stark2013}). Figure~\ref{StratIon}
shows that these ionization processes (boxes in figure) do appear with
different efficiencies in different parts of the atmosphere,
suggesting a brown dwarf atmosphere to be a stratified ionized medium
rather than a cold, neutral gas.

\bibliographystyle{spbasic}      
\bibliography{bib,bibChH,bib_slc}

%
%

\end{document}